**Multi-Stage Phase-Segregation of Mixed Halide Perovskites under Illumination: A Quantitative Comparison of Experimental Observations and Thermodynamic Models.**

*Klara Suchan, Justus Just, Pascal Becker, Carolin Rehermann, Aboma Merdasa, Roland Mainz, Ivan G. Scheblykin\* and Eva Unger\**

K. Suchan, I.G. Scheblykin, E. Unger

Division of Chemical Physics and NanoLund, Lund University, Box 124, 22100 Lund, Sweden

E-mail: Ivan.scheblykin@chemphys.lu.se, eva.unger@chemphys.lu.se

J. Just

MAX IV Laboratory, Lund University, PO Box 118, SE-22100 Lund, Sweden

P. Becker

Helmholtz-Zentrum Berlin für Materialien und Energie GmbH, Structure and Dynamics of Energy Materials, Hahn-Meitner-Platz 1, 14109 Berlin, Germany

P. Becker, R. Mainz

Helmholtz-Zentrum Berlin für Materialien und Energie GmbH, Microstructure and Residual Stress Analysis, Albert-Einstein-Straße 15, 12489 Berlin, Germany

C. Rehermann, A. Merdasa, E. Unger

Helmholtz-Zentrum Berlin für Materialien und Energie GmbH, Department Solution-Processing of Hybrid Materials and Devices, Keku léstraße 5, 12489 Berlin, Germany

eva.unger@chemphys.lu.se

C. Rehermann, E. Unger

Humboldt Universität zu Berlin, IRIS Adlershof, Research Group Hybrid Materials: Formation and Scaling, Am Großen Windkanal 2, 12489 Berlin, Germany

**ABSTRACT**

Photo- and charge-carrier induced ion migration is a major challenge when utilizing metal halide perovskite semiconductors for optoelectronic applications. For mixed iodide/bromide perovskites, the compositional instability due to light- or electrical bias induced phase-segregation restricts the exploitation of the entire bandgap range. Previous experimental and theoretical work suggests that excited states or charge-carriers trigger the process but the exact mechanism is still under debate. To identify the mechanism and cause of light-induced phase-segregation phenomena we investigate the full compositional range of methylammonium lead bromide/iodide samples, MAPb(Br$_x$I$_{1-x}$)$_3$ with x = 0…1, by simultaneous in-situ X-ray diffraction and photoluminescence spectroscopy during illumination. The quantitative comparison of composition-dependent in-situ XRD and PL shows that at excitation densities of 1 sun, only the initial stage of photo-segregation can be rationalized with the previously established thermodynamic models. However, we observe a progression of the phase-segregation that can only be rationalized by considering long-lived accumulative photo-induced material alterations. We suggest that (additional) photo-induced defects, possibly halide vacancies and interstitials, need to be considered to fully rationalize light-induced phase-segregation and anticipate our findings to provide crucial insight for the development of more sophisticated models.

## 1. INTRODUCTION

The seamless bandgap tunability of metal halide perovskites (MHPs) with the structure ABX$_3$ makes these semi-conductors an interesting absorber and emitter material for opto-electronic devices like solar cells and LEDs. Compositional stability is a prerequisite for both for the colour stability of LED devices, as well as for stable, reliable, and highly efficient solar cells. Halide mixing/alloying to X = (I,Br) or X = (Br,Cl) has been a common strategy to tune the MHP bandgap. However, this strategy is limited by the light-induced formation of low-bandgap phases for samples with bromide fractions of x > 0.17.[1–9] Here, the emission energy is observed to red-shift during illumination due to charge-carrier funnelling into the new low bandgap regions.[1] In LED devices this leads to colour instability of the emitted light. In solar cells with A= MA, MAFA or MAFACs with x>0.3, the device efficiency has been shown to drop significantly likely a result of charge-carrier trapping in I-rich domains and fermi-level pinning.[10–13] An in-depth understanding of the photo-segregation mechanism and how the process is affected by the bromide/iodide ratio is therefore of paramount importance to evaluate the operational stability and performance limits of MHP-based opto-electronic devices. In this work, we analyze the compositional dependence of photo-segregation to gain new insights into the role of local bandgap variations and inhomogeneous charge-carrier distributions in photo-segregation.

Numerous photoluminescence (PL) measurements on $MAPb(I_{1-x}Br_x)_3$, showed that the PL peak emission red-shifts upon illumination to approximately 1.7 eV, equivalent to a $MAPb(I_{1-x}Br_x)_3$ composition between x=0.15-0.2, for all compositions with x > 0.17.[1,2,2–7,9,14,4] Even for different cations and mixtures thereof, low-energy PL peaks corresponding to similar compositions have been shown to evolve.[3] The similarity in the peak energy of evolving PL spectra irrespective of the sample's composition, led to the hypothesis of the existence of thermodynamically-favored compositions adopted by the material upon phase-segregation[1,4,15–19]. This gave rise to the notion, that the photo-segregation process is nearly independent of the sample composition.[20,21] However, this notion is based on PL data, which is prone to probe the lowest accessible low energy emissive states in the sample.

X-ray diffraction (XRD) or optical absorption measurements have been used to investigate the compositional changes in mixed halide samples upon illumination.[1–3,14,22–28] These studies have, however, often only been carried out on samples in a narrow compositional range, with a focus on compositions between x = 0.4 and x = 0.6. Early XRD measurements showed the establishment of an I-rich minority phase (x = 0.2) and a Br-rich majority (x = 0.7) phase under illumination. Subsequent measurements failed to reproduce the result.[1,14,23–27] So far no agreement on evolution of structural properties upon illumination has been reached.[3]

While light-[1–3,6,7,22], bias-[27,29] and electron-beam induced phase-segregation[30,31] are shown to result from ion migration, highlighting the remarkable ion mobility in MHPs, the exact mechanism causing the ions to segregate is still under debate. Due to the reversibility of photo-segregation, it has been postulated to be a charge-carrier-induced transient state. [1–3] Brennen et al. compared and contrasted the models and categorized them into three categories:[3] The first category is based on the *generation of an electric field* induced by charge-carrier segregation or *a gradient in the generation rate* causing ions to drift.[32,33] For these effects to cause phase-segregation, a difference in the ionic mobility of bromide and iodide is a prerequisite. A second category of models postulated the *formation of polarons*[15,17,38] inducing local lattice strain and attracting or repelling certain charge-carriers. While some experimental observation could be rationalized, these models fall short e.g. in rationalizing the reported excitation density threshold of the photo-segregation.[7,39] This is accounted for in a third class of models, referred to as *charge carrier localization models*, triggered by the energy gain of photo-carriers when funneling into spontaneously formed low-bandgap domains.[7,19,40] What is yet not clear, and what will be further discussed herein, is how charge carrier localization leads to subsequent stabilization and growth of low-energy domains and how this causes changes in the phase-distribution in the entire sample.

To assess the validity of different models derived to describe the photo-segregation, this study investigated the steady-state phase-distribution of MAPb(Br$_x$I$_{1-x}$)$_3$ samples under illumination. For the *polaron model*, it was postulated, that mixed bromide/iodide samples should favorably segregate into phase-distributions dominantly centered around the compositions of x = 0.2 and x = 0.8 when illuminated, which would be consistent with the observation of an emissive state during PL measurements around x = 0.2.[15,17,18,41] The *charge-carrier localization model*, however, does not predict thermodynamically-preferred compositions despite the pure perovskites but predicts the favorable formation of low-energy states through which charge-carriers recombine as this stabilizes the excited state. This model hence contradicts the hypothesis, that the mixed bromide/iodide perovskite should be prone to segregate into specific phases e.g. with x = 0.8 and x = 0.2 independent of the bulk material composition.[41] What has to be noted is that these thermodynamic calculations only consider the initial stage of phase-segregation in which initial I-rich nano-domains are formed. In contrast to this, experiments show substantial fractions of the material in a phase-segregated state.[1,7,22]

We therefore set out to test existing hypotheses and models to see whether they are sufficient to describe all stages of phase-segregation observed experimentally. To probe both, the evolution of low-energy emitting sites and the overall phase-distribution of the samples, we studied the photo-induced phase-segregation in MAPb(Br$_x$I$_{1-x}$)$_3$ samples correlatively with PL and XRD across the entire compositional range. For this, we built a setup, shown in **Figure 1**, allowing us to keep samples in an inert environment, with a blue LED light-source for illumination at an equivalent of 1 sun enabling the simultaneous measurements of X-ray diffraction (XRD) and Photoluminescence (PL). This provided a dataset that allows us to compare the equilibrium phase-distribution for MAPb(Br$_x$I$_{1-x}$)$_3$ samples in the dark and light quantitatively and study the phase-segregation kinetics as a function of composition.

**Figure 1.** a) Schematic drawing of the atmosphere and temperature controlled measurement set-up developed to study the change in the structural and optical properties of MAPb(Br$_x$I$_{1-x}$)$_3$ simultaneously during illumination. b) PL spectra of a sample with $\bar{x}$ = 0.6 during illumination. The color represents the luminescence intensity. c) XRD patterns of the 200

reflex of the same sample during illumination, here the color represents the scattering intensity. Please note the log scale of the time and intensity.

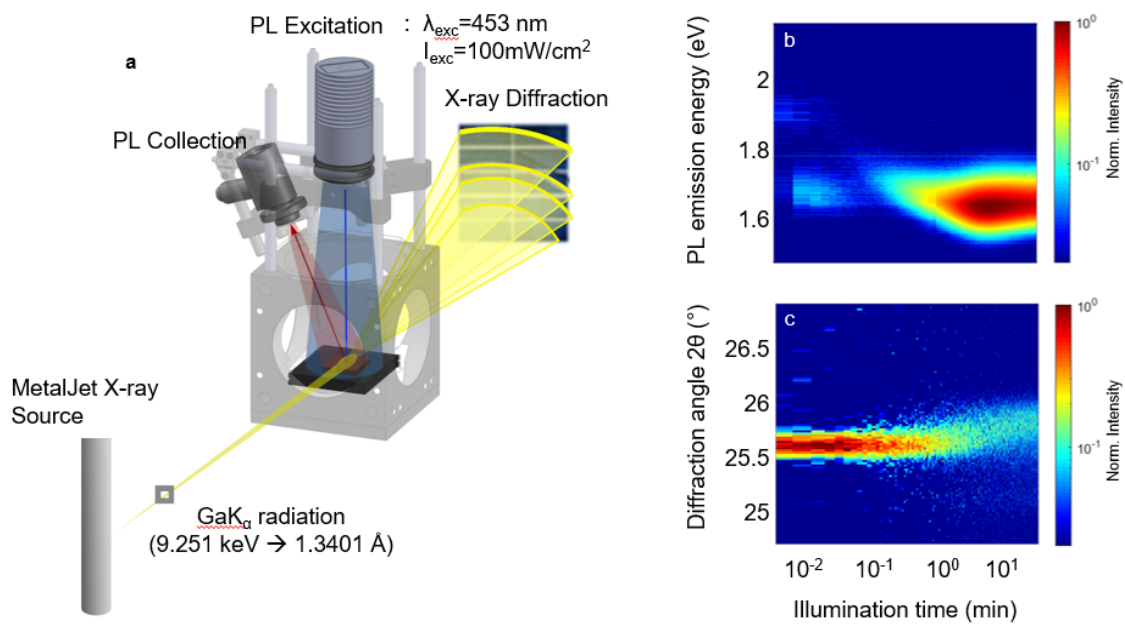

We observe that X-ray diffraction patterns adopted in equilibrium under illumination are highly dependent on the sample composition. The phase-distribution appears to be asymmetric and compositions with opposite bromide:iodide are not *mirror-images*. This indicates that the photo-segregation process does not affect bromide and iodide in the same manner. We derive and quantify the distribution of compositional phases as the *degree of segregation* and evaluate the quantitative evolution of the phase-distribution change through this parameter, *in-situ,* during phase-segregation. The comparison between experimental data and thermodynamic predictions yields, that at moderate charge-carrier densities, the average charge-carrier energy is not sufficient to rationalize the magnitude of changes in the phase-distributions observed. To reconcile the discrepancy between the thermodynamic models and the experimentally observed excited equilibrium states, we postulate that light-induced effects in the material are *accumulative*: phase-distribution changes are the consequence of more persistent changes in the material. In line with the experimentally observe apparent higher mobility of iodide compared to bromide ions, we postulate that preferential iodide-oxidation and lead-iodide bond cleavage causing iodide vacancies and interstitials,[34–37] is an effect that becomes greatly enhanced, when charge-carriers accumulated on iodide-rich domains causing dramatically-enhanced local charge carrier densities. We propose, that in the derivation of more refined theoretical models, both, thermodynamic driving forces as well as the effect of accumulative photo-induced changes in the material need to be considered.

## 2 Results

### 2.1 Equilibrium states

In **Figure 2**a, PL spectra and XRD pattern of the initial and final state for the samples of MAPb(Br$_x$I$_{1-x}$)$_3$ with $\bar{x}$ = 0, 0.1, 0.2, 0.4, 0.6, 0.8, 0.9, 1 are compared. The initial PL spectra are averaged over the first 100 ms of the measurement. The final PL spectra and XRD pattern are taken when both PL and XRD no longer change.

The initial PL spectra, as well as XRD patterns in the dark are consistent with literature data from the expected compositions[21,42,43] (see Figure S8). Further the initial broadening of the XRD patterns is at the lower end of previously reported peaks widths. (Figure S18) We therefore conclude that the average sample composition $\bar{x}$ is similar to the halide-ratio of the precursor solution, meaning that no halide is lost during preparation. Further we conclude that in the very beginning of illumination, samples are in a mixed state. For the sample with $\bar{x}$ = 0.6, the PL peak shifts so quickly, that within the first 100 ms, traces of the low PL peak are already visible in the spectrum.

Upon illumination, the steady state PL peak position shifts to lower energy. After prolonged illumination it is comparable for all mixed halide samples with $\bar{x}$ > 0.1[1,4] with a slight shift towards higher energy with increasing Br content. Hence irrespective of the sample's average composition $\bar{x}$, the PL emission occurs from states with a high relative iodide content of $x$ = 0.1 - 0.2. This similarity of the PL peak emission energy has been previously attributed to the formation of low-bandgap, iodide-rich, domains whose composition is largely independent of the bulk composition.[1,2,7,15]

Simultaniously, the XRD peak intensity decreases, and the peaks broaden and shift for all mixed halide samples as apparent from the normalized XRD patterns for the dark and illuminated equilibrium states, $I_{XRD_{ini}}(2\theta)$ and $I_{XRD_{fin}}(2\theta)$, for each $\bar{x}$. In contrast to the PL spectra, XRD-patterns of illuminated samples are characteristically different for samples of different average composition, $\bar{x}$. Pure samples, $\bar{x}$ = 0 and $\bar{x}$ = 1, exhibit no significant change. Samples with $\bar{x}$ < 0.5 exhibit broadened XRD peaks upon illumination whereas samples with $\bar{x}$ > 0.5 exhibit a shift of the most pronounced peak towards higher angles and a long tail towards lower angles. Our results can thus reconcile supposedly conflicting findings of previous reports[1–3,14,22–28] by demonstrating the strong dependence of the structural changes during photo-segregation on the average sample composition.

The long tail towards low diffraction angles, especially visible in the differential XRD pattern (SIX), is only visible for Br-rich compositions. I-rich samples do not exhibit a corresponding tail to higher angles. This asymmetry in the compositional dependence of the XRD patterns of the final equilibrium state under illumination becomes even more evident when translating the XRD-pattern to the compositional phase-distribution discussed in the next section.

**Figure 2**. a) PL spectra of the mixed state (grey) and the segregated equilibrium state under illumination with 1 Sun (color) for MAPb(Br$_x$I$_{1-x}$)$_3$ films with different sample composition $\bar{x}$. The PL of the photo-segregated film is independent of $\bar{x}$. b) X-ray diffraction patterns of the 200 (cubic) or 220/004 (tetragonal) reflex for different $\bar{x}$ (X-Ray wavelength 1.3401 Å) reveals the compositional dependence of the equilibrium segregated state under illumination. c) Shows that the initial and final position of the center of mass of the patterns remain constant during photo-segregation while a significant shift in the position of the majority phase occurs for Br-rich samples.

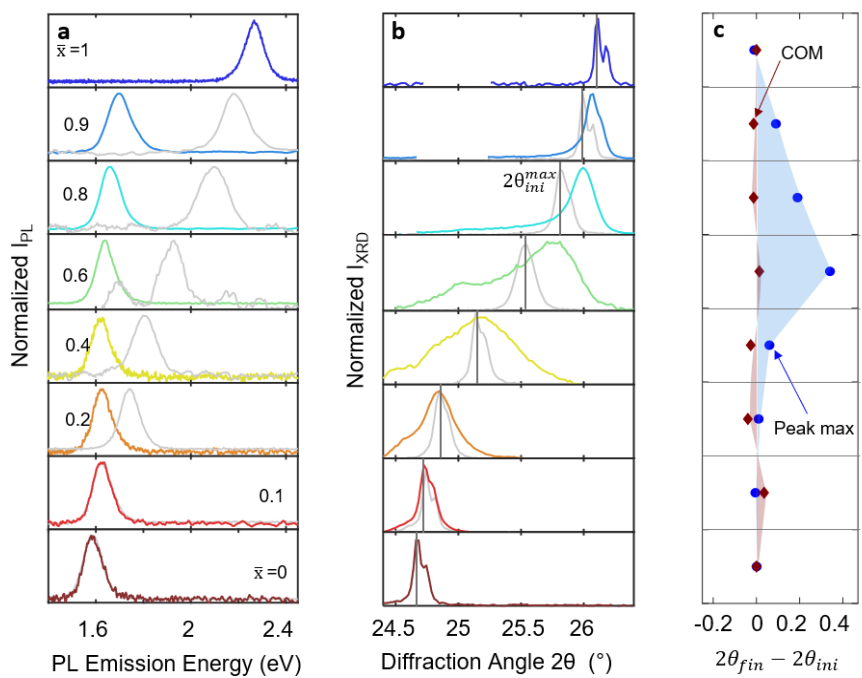

From the XRD patterns as well as the PL peak positions, information about the compositional distribution as well as composition of the emissive state can be derived as discussed in section 5. Here we assume, that the observed changes in the XRD pattern (broadening, shift and appearing of a tail) are primarily caused by photo-segregation of the samples into domains with differing local compositions. We believe that this is a valid approach as we were not able to reconcile the observed shift, the evolution of a tail and broadening of the experimental XRD patterns in conjunction with their constant COM with changes in strain or crystal domain sizes, (see SI Note 3 and 6). We cannot, however, rule out an additional contribution of size effects especially for small I-rich domains.

The normalized compositional phase-distributions in the dark $\Phi_{ini}(x)$ (empty bars) and after an equilibrium distribution has been reached under illumination $\Phi_{fin}(x)$ (colour) are shown in **Figure 3**a together with the composition of the emissive phase upon illumination (yellow). $\Phi_{ini}(x)$ is rather narrow and centered around $\bar{x}$ for all samples. Upon illumination a broad distribution of compositional phases forms. Comparing $\Phi_{ini}(x)$ and $\Phi_{fin}(x)$ for all compositions, it becomes obvious, that the equilibrium distribution under illumination depends on the sample composition $\bar{x}$.

In I-rich samples with $\bar{x} < 0.5$, $\Phi_{fin}(x)$ is fairly symmetrically with respect to $\Phi_{ini}(x)$, meaning that we mostly observe a broadening of the phase-distribution. States that are slightly richer in I and Br form almost equally. Contrarily, in Br-rich samples up to 70% of the materials changes its composition. Furthermore, for $\bar{x} > 0.5$ the phase-distribution under illumination appears much more asymmetric compared to the mixed state in the dark. Especially for the samples with $\bar{x} = 0.8$ and $\bar{x} = 0.9$, the majority of the sample adopts a narrow phase-distribution with the maximum centred at higher Br-content. Additionally, a very broad tail towards I-enriched compositions forms. In contrast, I-rich samples, such as $\bar{x} = 0.2$, do not exhibit such a tail neither towards higher Br-content nor towards higher I-content. This becomes even clearer when comparing the $\Phi_{fin}(x)$ for samples of the "mirror" compositions ($\bar{x} = a = 0.2$ and $\bar{x} = 1 - a = 0.8$) in Figure 3b, showing that these phase-distributions are not "mirror images" of each other. This highlights a principal difference between I-rich and Br-rich samples indicating that iodide and bromide ions are affected differently during the photo-segregation process. The shift of the maximum of the phase-distribution towards higher bromide content and evolution of a long tail distribution of iodide-richer phases can be interpreted as a favored migration of iodide ions out of mixed halide domains causing the parent domain to become iodide depleted and bromide enriched. The migration of iodide into adjacent domains causes a gradient of iodide within the sample leading to the long compositional tail towards higher iodide-content domains.

What we would like to emphasize is that we do not observe the adoption of any preferred compositions comparing samples of different $\bar{x}$ upon illumination. Viewing both XRD and PL data translated to composition as shown in Figure 3a highlights an important fact: While PL measurements allude to the composition of the emissive states, $x_{PL}$, to be fairly similar for all samples and hence independent of the sample's average composition, XRD measurements reveal that the compositional distribution $\Phi(x)$ differs strongly between samples. The volume fraction of $x_{PL}$ within $\Phi_{fin}(x)$ becomes smaller with increasing average Br-content. Hence, $x_{PL}$ does hence not reflect the composition of a thermodynamically preferred phase but rather the phase with the highest fraction of iodide within the sample. For Br-rich samples, the relative volume of these domains is so small that it is hardly detectable by XRD. For the $\bar{x} = 0.6$ sample, $x_{PL} \approx 0.1$, which means that the entire PL originates from domains with only 10% bromide that constitute less than 1% of the total sample volume. Thus, the charge-carriers predominantly generated around $x_{MAX} = 0.75$ funnel to only about 1% of the material with $x_{PL} \approx 0.1$.

This charge-carrier funneling leads to an increased charge-carrier concentration within the I-rich domains emission is stemming from. The relative charge-carrier density in the emissive phase $n_{x_{PL}}$ can be estimated in dependence of its volume fraction $\Phi_{x_{PL}}$, as $n_{x_{PL}} \propto 1/\Phi_{x_{PL}}$. Because $\Phi_{x_{PL}}$ is decreasing severely with $\bar{x}$, $n_{x_{PL}}$ follows this dependency. A rough estimate of the relative charge-carrier density is obtained assuming, that nearly all charge-carriers funnel to the I-rich domains with composition $x_{PL}$. This is a justified assumption, as the high energy band in PL is quenched entirely during photo-segregation, signifying a depletion of charge-carriers in the Br-rich domains. In a sample with $\bar{x} = 0.2$, $\Phi_{x_{PL}}^{fin} \approx 0.1$ such that the local charge-carrier density in the emissive phase increases by $1/\Phi_{x_{PL}}^{fin} \approx 10$. In a sample with $\bar{x} = 0.8$, where, $\Phi_{x_{PL}} \approx 0.01$ similar domains exhibit a 100 fold increased charge-carrier density. In a segregated sample with $\bar{x} = 0.8$, an illumination with 1 sun, thus results in a local charge-carrier density within the emissive phase, equivalent to illumination with 100 suns. It must be noted that the volume fraction of the emissive phase growths during segregation as discussed in detail in section 2.2. In the initial stage of phase segregation, the volume fraction is below the detection limit of XRD and thus the charge-carrier densities are likely to be many times larger than estimated for the final segregated state.

In the case of $\bar{x} = 0.1, 0.2\ and\ 0.4$, the XRD data indicates that phases with even higher I-content than $x_{PL}$ form. We cannot rule out, that this is an artefact, due to domain-size related, or phase-transition related broadening/shift of the XRD peaks. If samples indeed exhibit more iodide-rich domains than $x_{PL} \approx 0.1$, we do not observe luminescence from them, which might indicate that they are very defective and hence dark.

As the evolution of the phase-distribution depends significantly on the sample's average composition $\bar{x}$ we introduced a quantitative metric to compare the degree of segregation between samples of different composition, $D_{seg}$, as defined in section 5, equation 7. $D_{seg}$ measures how much the compositional distribution has changed compared to the initial state. This is reflected in $D_{seg}(\bar{x})$ which is proportional to the Br-content (Figure 3c). We do not observe an instability maximum, but rather the instability increases up to $\bar{x} = 0.9$. This trend agrees with the increased difference between the majority phase and the emissive phase $x_{MAX} - x_{PL}$ for higher Br contents.

**Figure 3.** a) Distributions of local compositions $\Phi(x)$ in the mixed film (empty bars) as well as in the equilibrium state under prolonged illumination (color) are calculated from XRD (See SI) and compared to the composition of the emissive phase $x_{PL}$ (yellow) calculated from the final PL peak position. In Br-rich samples, is the emission coming from utmost I-rich domains making up a minuscule portion of the sample. b) A comparison of the compositional distribution of the mixed and segregated films for $\bar{x}=0.8$ and $\bar{x}=0.2$ contrasting the symmetrical broadening of I-rich samples with the asymmetric changes in the compositional distribution for Br-rich samples. c) The degree of segregation $D_{seg}(x)$ increases with increasing bromide content in parallel to the increasing difference in composition between the majority phase $x_{MAX}$ and the emissive phase $x_{PL}$.

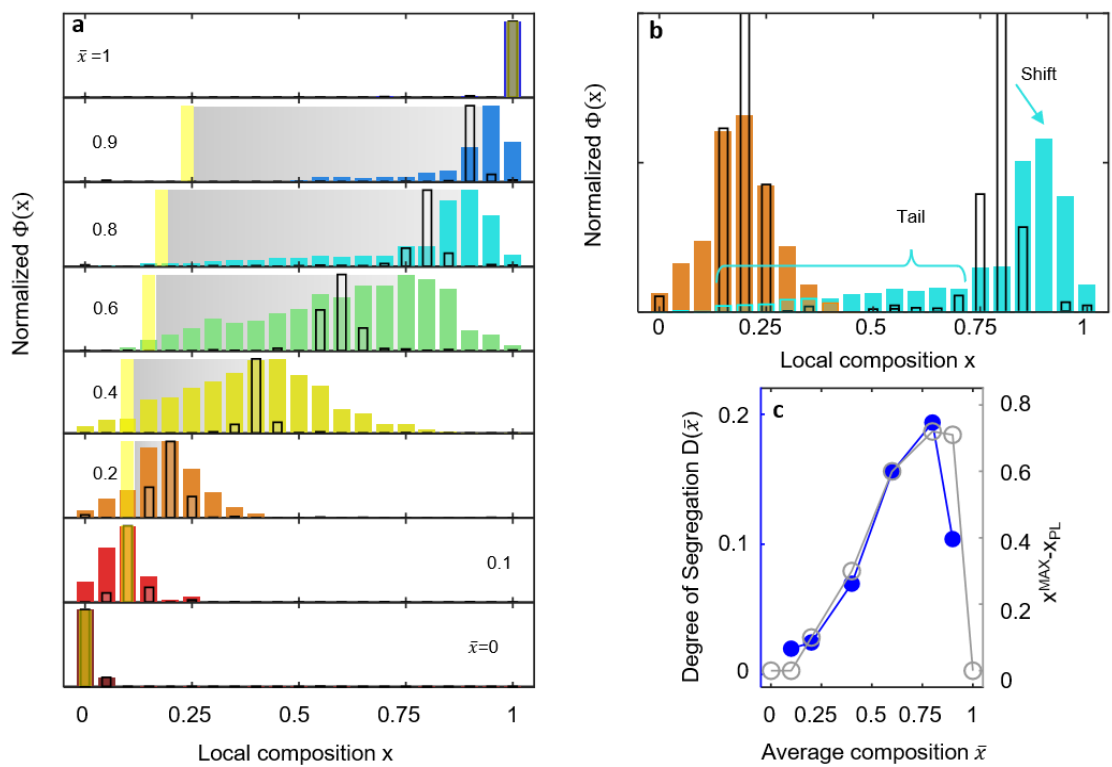

## 2.2 Kinetics

The real-time simultaneous in-situ measurements of XRD and PL allow us to directly compare and contrast the evolution of the structural changes occurring during phase segregation and the corresponding changes in the emission. The distribution of charge-carriers within the film is responding to changes in the compositional distribution, by localization to emerging low bandgap domains. As the structural changes and the charge-carrier distribution are interdependent, a simultaneous analysis of the evolution of the phase-distribution as well as the low-energy emitting sites can further the understanding of the photo-segregation process.

The evolution of the degree of segregation, $D_{seg}(\bar{x}, t)$, which gives an average measure of how much the sample has changed with respect to the initial dark equilibrium state as a function of time $t$, allows to directly track how the new phase equilibrium is established under illumination (solid, colored lines **Figure 4**a). The experimental evolution of $D_{seg}(\bar{x}, t)$ follows a simple exponential approximation for all investigated compositions as expected for a first order process. Comparing samples with different average composition $\bar{x}$, the phase-equilibrium under illumination is established faster for samples with higher bromide content. For a quantitative comparison, we calculated the segregation rate $k_{seg}$ defined as the inverse time needed for $D_{seg}(\bar{x}, t)$ to reach $1/(1-e) \approx 0.63$ of its maximum value (Figure 4c). We observe that $k_{seg}$ increases with Br-content. For example, the sample with $\bar{x} = 0.9$ segregates almost 6 times faster than the "mirror-composition" sample with $\bar{x} = 0.1$ with rates 0.45 min$^{-1}$ *vs* 0.08 min$^{-1}$ respectively.

The volume-fraction increase in iodide-rich domains can be quantitatively followed by analyzing the XRD signal in the angular range expected for the composition of low-energy emission sites $I_{XRD}(2\theta)(x = x_{PL}, t)$ (dotted coloured lines Figure 34) In iodide-rich samples, the growth of iodide-rich domains, $I_{XRD}(2\theta)(x = x_{PL}, t)$, correlates with the overall phase-segregation, measured through $D_{seg}(t)$, while in Br-rich samples the growth of domains with a composition of $x = x_{PL}$ is lagging behind the overall phase-segregation. This indicates that in Br-rich samples, the majority of the sample first segregates into bromide and iodide enriched phases close to the mixed composition and in a second step the I-rich tail is formed.

Zooming into the first seconds of photo-segregation, the previously noted complex evolution of PL becomes apparent (Figure 4b).[44] A fast decrease of the high energy emission at 1.93 eV by two orders of magnitude can be seen (stage 1). This decrease in high energy emission is accompanied by the appearance of a low energy PL-peak, which disappears after a few seconds, as we reported previously.[44] The energy of the PL emission can be taken as a measure of which phase is populated with charge-carriers, if band-to-band emission dominant. The fast shift from high-energy PL to low energy PL in stage 1, thus signifies the charge-carrier depletion of Br-rich domains and the charge-carrier funneling to and localization in emerging low energy domains.

As a first approximation, the fraction of the low energy PL ($I_{PL_{low}}/(I_{PL_{low}} + I_{PL_{high}})$) can be used to follow the extend of charge carrier funnelling (grey traces in Figure 4a). In comparison to the evolution of $D_{seg}(t)$ and $I_{XRD}(2\theta)(x = x_{PL}, t)$, it becomes evident, that for all samples with $\bar{x} > 0.2$ the charge-carrier localization to I-rich domains occurs rapidly already within the first seconds of phase segregation, prior to any major I/Br ionic reorganisation and the growth of a substantial low-bandgap phase in stage 2 (Figure 4a).

The comparison of the kinetics of PL and $D_{seg}(t)$ for different timeframes from seconds to minutes, therefore shows, that the segregation process is a 2-step process. We conclude that the nucleation and population of nanoscopic I-rich domains, too small to be detectable in XRD is occurring in stage 1, preluding the phase segregation of the bulk material in stage 2.

In numerous studies, the phase-segregation kinetics have been analyzed based on the evolution of the PL signal intensity $I_{PL}(t)$.[1–3,6,7,15,22,23,32,39,44,45] The PL intensity is determined by the absorption of the material and ratio between radiative and non-radiative recombination kinetics. An increase in emission signal thus signifies a decrease of non-radiative recombination relative to radiative recombination, caused by phase-segregation[7]. The PL intensity is however not directly connected to the volume-fraction of iodide-rich domains. In the SI we show that there is indeed a rough correlation between the PL-yield and the degree of phase segregation (Figure S14). However, this surprising correspondence is only valid for specific, low-bromide, sample compositions. Our observation affirm that experimental works utilizing the PL intensity evolution as a probe for the phase-segregation kinetics [1–3,6,7,15,23,28,32,39,44,45] derive roughly valid phase-segregation rates from their experimental data especially if measurements were carried out on samples with a high-iodide content. PL is however not a reliable probe for the phase-segregation kinetics for samples with higher bromide content and should hence not be used as a quantitative probe to study the phase-segregation kinetics.

**Figure 4.** a) The kinetics of the degree of segregation (solid colored line), the XRD intensity of the tail states, corresponding to $x_{PL}$ (colored dotted lines) and the fraction of low energy PL intensity $\frac{I_{PL_{low}}}{I_{PL_{low}}+I_{PL_{high}}}$ (grey) are compared for compositions $\bar{x}$ =0.2 to 0.9. b) Zoom into the initial phase of photo-segregation. Comparing the evolution of the high energy PL (grey - thin) the low energy PL (grey - bold), the fraction of low energy PL (red) and the segregation kinetics (green) for the sample with $\bar{x} = 0.6$ shows, that a very fast initial process is occurring in the first seconds corresponding to a rapid charge-carrier depletion of the Br-rich phase, while the macroscopic phase segregation occurs significantly slower at the order of minutes.

c) The segregation rates of the slow macroscopic phase segregation increase significantly for Br-rich samples.

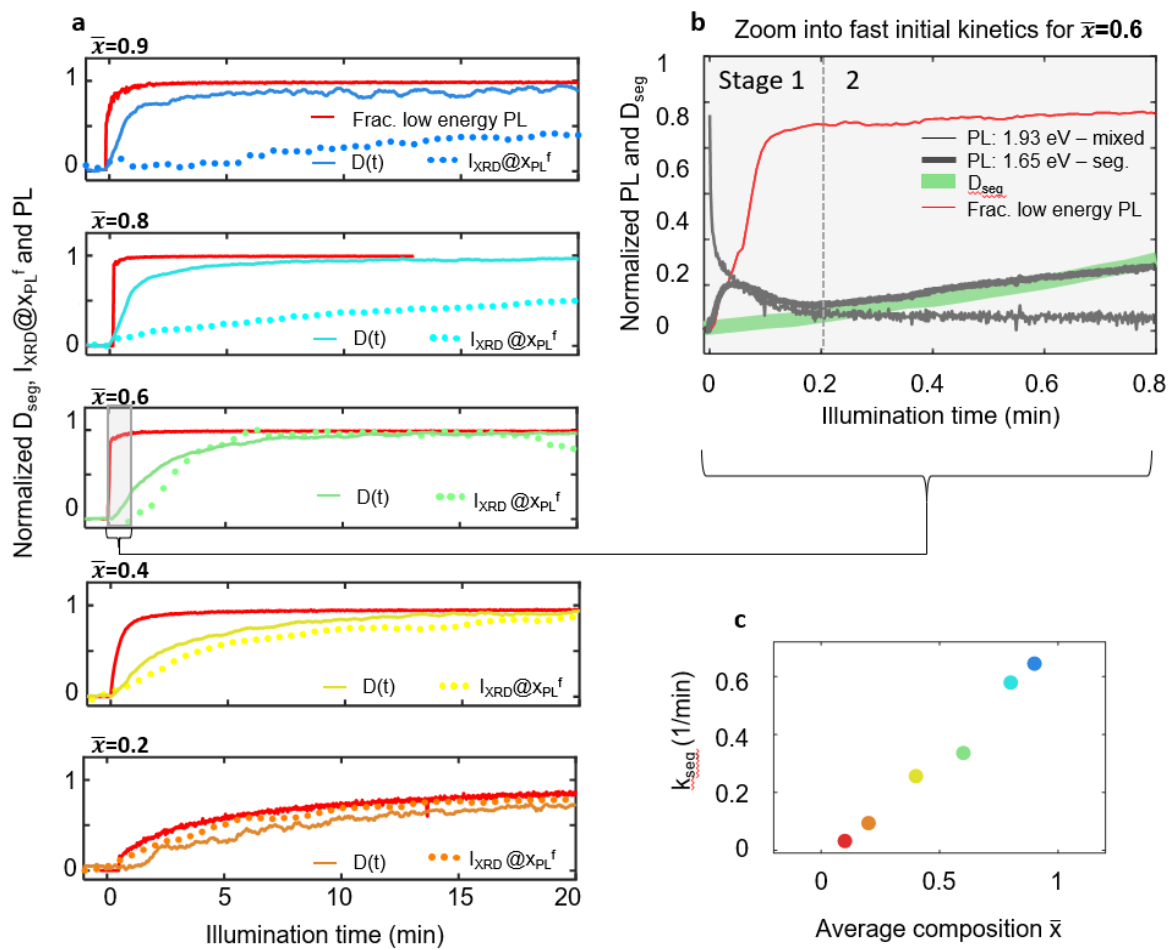

## 3. Discussion

The cause of photo-segregation has been postulated to be a charge-carrier-induced transient state. The exact mechanism however, how the excitation of charge-carriers leads to a redistribution of the halide ions within the crystal remains subject of debate. Combining both PL and XRD allows us to directly investigate the interplay of inhomogeneous charge-carrier distribution and structural/compositional rearrangement.

Our experimental results show that, evidently, the photo-induced phase-segregation is an asymmetric process with respect to the halide ratio. There is no mirror symmetry in I-rich and Br-rich samples. This is particularly apparent when comparing the excited state phase-distribution of compositions with mirror bromide/iodide ratios (e.g. $\bar{x} = 0.2$ and $\bar{x} = 1-0.2=0.8$), where the phase-distributions of I- and Br-rich samples in the illuminated state do not mirror each other. While in I-rich samples the phase-distribution merely broadens, in Br-rich samples photo-segregation leads to a change in composition in up to 70% of the material and a broad distribution of local compositions establishes. Furthermore, the segregation kinetics depend on the sample's average composition exhibiting faster phase-segregation kinetics for samples with higher bromide content. This shows that all aspects of the photo-segregation process, including compositional stability, segregation kinetics, which compositional phases form as well as how much of the material undergoes photo-segregation, depends on the average sample composition. Further we observe two distinct stages in the segregation kinetics. A temporary low energy PL peak appears rapidly upon illumination (Figure 4b stage 1), as discussed in detail in our previous work.[44] At this fast timescale no change is observable in the XRD patterns yet. The much slower segregation kinetics visible in XRD on the timescale of several minutes and the corresponding slow evolution of the final low energy PL Peak, show that substantial changes in the halide distribution occur only in the second stage, after the formation of I-rich nano-domains and the localization of carriers there (Figure 4b). This highlights the value of using PL and XRD as complementary probes, as PL is very sensitive to the appearance of even miniature I-rich domains, while XRD gives an accurate picture of the overall compositional phase-distribution in the sample.

### 3.1 Qualitatively, charge-carrier localization causes photo-segregation process

Using these experimentally obtained phase-distributions in the segregated state in combination with the segregation kinetics allows us to validate whether previously developed thermodynamic models can describe the observed segregation process. Currently, the predominant theoretical models to describe photo-segregation are the *polaron* model [15,17,18,41] and a model based on the *localization of charge-carriers* in low bandgap domains.[7,19,40] Both models assess the compositional photo-stability of Br-I perovskites by estimating whether or not the nucleation of phase-segregation is energetically favorable. Nucleation of phase-segregation here means formation of miniscule I-rich or Br-rich domains as the onset of phase segregation. We assign this nucleation of phase-segregation to the fast initial appearance of low energy PL in stage 1 (see Figure 4b). Often only this initial stage is considered theoretically, without describing the consecutive change of the majority of the material occurring later in time. Further, the models often assume an idealized segregated state consisting of only two phases, a Br-rich phase and an I-rich phase.[7,15,40] As we obviously observe broad and composition-dependent phase-distributions in illuminated samples, we found it necessary expand the theoretical models.

The free energy of mixing is used as a measure for the sample's compositional stability in dependence of the average sample composition. The free energy of mixing $\Delta_{mix}F(\bar{x})$, is the free energy gain when mixing the pure I and Br phases to form the mixed phase with composition $\bar{x}$, with respective free energies, $F^I$, $F^{Br}$, and $F^{mix}$. A first approximation of this yields: $\Delta_{mix}F(\bar{x}) = F^{mix} - (\bar{x}F^{Br} + (1-\bar{x})F^I)$. In the excited state, a charge-carrier term has to be added to calculate the free energy of mixing $\Delta_{mix}F^*$. The nature of this charge-carrier term depends on the model used.[7,15,18,40]

In the *polaron model* developed by Ginsberg et al., the charge-carriers are assumed to strain the lattice resulting in an additional strain-related energy term, $\Delta g_s$, which can be reduced by halide segregation.[15] From the polaron model, the phase compositions of $\bar{x} = 0.2$ and $\bar{x} = 0.8$ are predicted to be energetically favorable and thus most stable under illumination. The predominant formation of these phases, independent of the sample's average composition is expected. The agreement between the theoretical predictions and the experimentally observed PL spectra, was seen as a validation of the polaron model.[15,17,18,41] However, our experimental results show that the compositions of the majority and minority phases strongly depend on the average sample composition $\bar{x}$. For $\bar{x} = 0.6$, we indeed observe a majority phase with $\bar{x} = 0.8$. However, for the sample with $\bar{x} = 0.9$, predicted to be photo-stable in the polaron model, we observe a majority phase, which resembles the pure Br-perovskite instead. We therefore cannot confirm the existence of generally favored compositions. Further, due to the theoretically predicted symmetry of the free energy of mixing and the minimum in free energy for $\bar{x} = 0.8$[15,18] the model fails to describe the asymmetry between I-rich and Br-rich sample's as well as the compositional instability of samples with $\bar{x} \geq 0.8$. We therefore conclude, that the *polaron model* in its current form cannot be employed to rationalize the full set of our experimental data.

In the *charge-carrier localization model*, derived by Kuno et al.[7,19] and Bobbert et al.[40], the formation of a certain amount of I-rich domains is favorable, as it allows the sample under illumination to reduce its free energy by localization of charge-carriers in the I-rich domains, were the bandgap $E_G$ is substantially smaller. In the mixed state the energy of the charge-carriers $N \cdot E_G(\bar{x})$, where N is the number of charge-carriers and $E_G(\bar{x})$ is the bandgap of the mixed material. As shown schematically for a sample with $\bar{x} = 0.8$ (**Figure 5** d,e), if the sample is in the dark, only the mixed phase exists with $x = 0.8$ with a free energy $F(\bar{x})$. Under illumination the free energy is increased by the energy of the charge-carriers $F^*(\bar{x}) = F(\bar{x}) + N \cdot E_G(\bar{x})$. The free energy of the illuminated state is thus a function of the number of excited states N as well as the bandgap of the material $E_G$. The increase in energy is thus higher for bromide-rich samples, which have a higher bandgap. However, it can be effectively reduced by charge-carrier funneling into pre-existing or randomly created small I-rich domains to $N \cdot E_G(x_{PL})$, where $x_{PL}$ is the composition of the emissive phase.[7,19,40] The difference in bandgap between the mixed phase and the I-rich phase therefore provides a thermodynamic driving force for the formation of I-rich domains. This model predicts the free energy of mixing under illumination to be bandgap dependent and thus asymmetric and almost linearly increasing with increasing Br-content. This agrees with the herein observed higher final degree of segregation, $D_{seg}$ with higher bromide-content, shown in Figure 3.

Considering the formation of an I-rich phase and charge-carrier funneling to these domains is enough to rationalize the experimentally observed compositional distribution established under illumination. In Br-rich samples, the majority phase is nearly completely deprived of charge-carriers. This phase is thus stabilized because it is virtually in the dark as all charge-carriers funnel to few low-bandgap domains. Which composition the majority phase adopts thus depends merely on the sample's average composition and on how much iodide is accumulated in the iodide rich phase, following $x_{Br} \approx \bar{x} - \Phi_{x_{PL}} \cdot x_{PL}$ with $\Phi_{x_{PL}}$ being the volume fraction of the iodide rich phase. This agrees with the experimentally observed composition-dependence of the majority distribution. A second energetic minimum to rationalize the formation of a bromide-rich phase is thus not needed: Because the majority distribution is virtually in the dark, it is striving to stay mixed. The width of the majority distribution is thus observed to be narrower in Br-rich samples than in I-rich samples. In I-rich samples, the charge-carrier distribution is expected to be more homogeneous. We thus conclude that the charge-carrier localization model is qualitatively describing the experimental observed asymmetry of the photo-segregation process and the observed compositional distribution very well.

**Figure 5.** In stage 1, nanoscale segregated domains are formed. Their formation is locally energetically favorable as indicated in the energy schematics (right side). The presence of I-rich domains (red ball) in close vicinity to charge-carrier generation allows for the reduction of charge-carrier energy via funneling to such low bandgap, I-rich domains. The bromide rich phase (green ball) is thus depleted of charge-carriers and is effectively in the dark. Triggered by the nucleation of I-rich domains, the sample undergoes macroscopic photo-segregation in stage 2, which is kinetically stabilized by a difference between fast segregation rates and slow recovery rates, as indicated in, as indicated in the energy diagram corresponding to the entire sample.

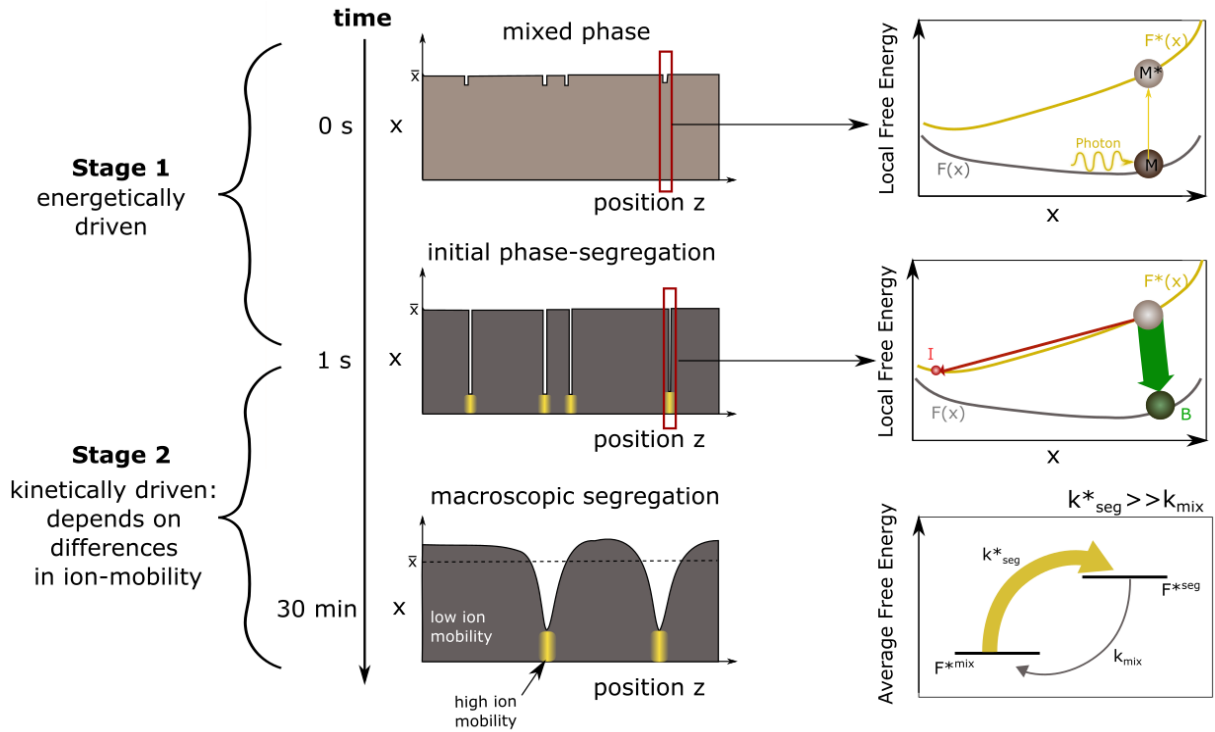

### 3.2 Quantitatively, the charge-carrier density is too low to segregate the entire sample

We further set out to assess whether the charge-carrier localization model can describe our experimental data quantitatively. In a static picture, one state is favored over the other if it is lower in energy. In the charge-carrier localization model this means that under illumination, the energy of the segregated state needs to be less than of the mixed state. We use the experimentally obtained the volume fractions $\varphi(x_m)$ to approximate $\Phi(x)$ and the PL energy to estimate the energy of the charge carriers. We can thus calculate this difference in free energy $\Delta E^*(\bar{x}) = F^*_{mix}(\bar{x}) - F^*_{seg}(\bar{x})$ directly, while relying on literature values for the free energy in the dark.

Distinct equations can be derived for the illuminated state free energy in the segregated state and the mixed state.

$$F^*_{mix}(\bar{x}) = \underbrace{F(\bar{x})}_{dark} + \underbrace{N \cdot E_G(\bar{x})}_{charge\ carrier\ term} \quad \text{(Equation 1)}$$

$$F^*_{seg}(x) = \underbrace{\int_0^1 \Phi^{seg}(x) \cdot F(x)dx}_{dark} + \underbrace{N \cdot E_G(x_{PL})}_{charge\ carrier\ term} \quad \text{(Equation 2)}$$

The number of charge-carriers N is proportional to the generation of charge-carriers and their lifetimes and is in first approximation not assumed to change during the course of phase-segregation.

The energy gained by segregation can therefore be expressed as:

$$\Delta E^*(\bar{x}) = F^*_{mix}(\bar{x}) - F^*_{seg}(x)$$

$$= \underbrace{F(\bar{x})}_{dark} + \underbrace{N \cdot E_G(\bar{x})}_{\substack{charge\ carrier\\mixed}} - \underbrace{\int_0^1 \Phi^{seg}(x) \cdot F(x)dx}_{dark} + \underbrace{N \cdot E_G(x_{PL})}_{\substack{charge\ carrier\\segregated}}$$

$$= \underbrace{F(\bar{x}) - \int_0^1 \Phi^{seg}(x) \cdot F(x)dx}_{dark} + \underbrace{N \cdot (E_G(\bar{x}) - E_G(x_{PL}))}_{charge\ carrier} \qquad \text{(Equation 3)}$$

In the dark, if $N = 0$, $\Delta E^*(x)_{N=0}$ is negative for all compositions, showing that, without the contribution of charge-carriers, the segregated state is energetically unfavorable. As expected, our estimation thus yields, that phase-segregated samples will recover in the dark, in line with the reversibility of the phase-segregation observed experimentally.[1]

The phase segregation may become energetically favorable if the charge-carrier contribution is large enough.

$$N(E_G(x_{PL}) - E_G(\bar{x})) > F(\bar{x}) - \int_0^1 \Phi^{seg}(x) \cdot F(x)dx \qquad \text{(Equation 4)}$$

As the charge-carrier term scales with N, a more detailed look at realistic charge-carrier densities is needed to assess whether segregation is predicted to be favorable under realistic conditions.

The thermodynamic models, predicting photo-segregation to be energetically favorable, do their predictions based on a simplified model system. Either very high charge-carrier densities have been assumed with *n* on the order of one charge-carrier per formula unit (f.u.)[7,18] (formula unit is MAPb(I,Br)$_3$ which has a volume of ≈0.25nm$^3$ )[40]. Or only the very initial stage of photo-segregation with $\Phi(x_{PL}) \approx 0$ and thus hardly any change in the compositional distribution is considered.[40] Gao et al. consider a polaronic strain of δ = 0.15, which is only plausible to occur very localized around each excitation. At reasonable excitation densities this does not reflect the average strain of the entire sample.[18]

With an excitation density of 100 mW·cm$^{-2}$ (~1 sun) at 453 nm, and a charge-carrier lifetime of 100 ns, which is reasonable metal halide perovskites[46], the average charge-carrier density is approximately $n = 1.5 * 10^{-6}\ nm^{-3}$ (which is $1.5 * 10^{15}\ cm^{-3} n\ or\ 4 * 10^{-7}\ /f.u.$). In a grain, 300nm·300nm·200nm, this corresponds to 25 charge-carriers only. The reduction in energy of these 25 charge-carriers needs to be enough to segregate the entire grain. The blue line in **Figure 6**, shows that under the experimental conditions, the contribution of the mean charge-carrier density is not sufficient to explain the extend of photo segregation, experimentally observed, where up to 70% of the material undergo a change of their composition. The segregated state only becomes stable under the assumption of unreasonable large charge-carrier lifetimes on the order of 1 ms (SI Note 11). We therefore conclude, that the experimentally observed equilibrated segregated state, is not energetically favorable over the mixed state at an excitation density corresponding to 1Sun.

However, as discussed in section 2.1, the charge-carrier funneling to very small domains creates locally very high charge-carrier densities. Such high charge carrier densities can rationalize the initial stage of phase-segregation, as predicted by theoretical modelling.[7,40,41] In Figure 6 the colored lines indicate the energy gained by segregation for the excited state for different charge-carrier concentrations $n_I$ within the I-rich domains. The segregated state only becomes locally energetically favourable assuming high local charge-carrier concentrations $n_I = 25,000 * n$. This is the case, if all charge-carriers generated in the exemplarily 300nm*300nm*200nm grain funnel to a 5nm*5nm*5nm cube.

Both, models based on the localization of charge-carriers as well as the polaron model, are static models, were charge-carriers only have a momentary effect, which is gone once the charge-carriers recombine. Each charge-carrier can only have a small effect due to their limited energy of ~2eV. At moderate charge-carrier densities, found for 1 sun, only a very small volume fraction of the material around $10^{-6}$ can therefore be rationalized to segregate. This is corresponding to the initial state of phase segregation and in line with the theoretically predicted phase segregated nuclei of ~1nm in size.[7] We propose that the initial rapid appearance of a low energy PL peak, prior to the macroscopic photo-segregation indeed corresponds to the appearance of such locally segregated domains, which are too small to be visible in XRD. As observed from the PL kinetics (Figure 4b), already after a few seconds the entire emission is stemming from I-rich domains. In this initial stage of phase segregation, miniscule I-rich domains allow for effective reduction of the charge-carrier energy but no substantial changes in the compositional distribution have yet occurred, such that the 'dark energy term' is virtually staying constant. While it becomes clear, that such a state will be thermodynamically preferred, it is not trivial to rationalize why the sample continuous to segregate over a time of 30 minutes leading to the majority of the sample to change its composition. We concluded that to explain the evolution of phase-segregation, other factors and phenomena needed to be considered.

**Figure 6**. Shows that in the dark the energy gained by segregation is negative for all compositions indicating that the segregated state is unstable and will remix in the absence of charge-carriers. For an excitation with λ=453nm equivalent to 1 sun (100mW/cm$^2$), $\tau$=100 ns and homogeneously distributed charge-carriers (turquoise) the charge carriers do not possess enough energy to segregate the sample. The energy gained by segregation for higher excitation

densities, which may locally be reached by different degrees of charge-carrier accumulations in iodide rich domains are shown in colored lines. At least 25 000x increase in charge-carrier density is needed to stabilize the segregated state.

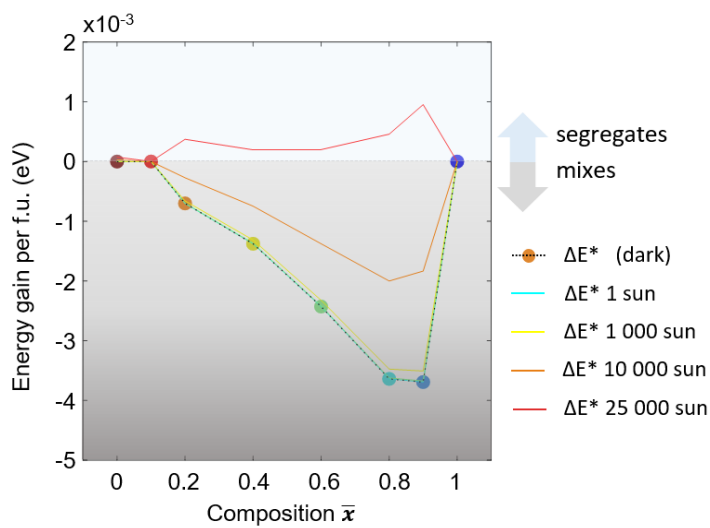

### 3.3 Charge-carrier localization causes accumulative changes in the material

To rationalize the extended phase-segregation towards the final phase segregated state under illumination, we propose a dynamic picture in which the accumulative energy exerted on the sample by continuous irradiation leads to long-lived changes to the material. Kuno et al. developed a phenomenological kinetic model in which it is pointed out, that in addition to an energetic driving force, the differences between the phase-segregation rate and recovery (remixing) rate determine the final state under illumination of a sample.[7] We here go beyond this and propose that a light induced spatial inhomogeneity of segregation and recovery rates, is needed to explain the latter stage of macroscopic photo-segregation.

We propose that this inhomogeneity is caused by a prevalent formation of photo-induced halide vacancies in I-rich domains resulting in spatially inhomogeneous ion mobilities. The role of halide-vacancies as accelerating factor for photo-segregation has recently been reviewed by Tian et al. .[47] main mechanism is, that ion-migration, which is the key pre-requisite for both segregation and recovery predominantly occurs via vacancy-mediated diffusion of halide ions.[35,48,49] [6,12] Next to pre-existing halide vacancy densities, photo-induced vacancies via weakening of Pb-I bonds need to be considered.[35–37,49] They are likely the major reason behind the reported excitation-density dependence of the ion-mobility[35,36,49–53] and have recently been suggested to be an underlying factor determining phase segregation.[34]

In the case of mixed halide perovskites, the inhomogeneous charge carrier density implies that photoinduced effects primarily occur in the I-rich domains. The concentration of charge-carriers in 'hot-spots', leads to a higher probability of oxidation and creation of halide vacancies there, than expected for irradiation with 1 sun. By continuous irradiation, local high iodide vacancy densities are thus built up, which results in locally enhanced ionic mobility. This inhomogeneous ion-mobility leads to large differences between $k_{seg}$ and $k_{mix}$. The phase segregation which is only occurring in the 'hot spots' with high ionic mobility will be very fast. On the contrary, the recovery which occurs in the charge-carrier deprived majority of the sample, where the ionic mobility is expected to be much smaller, will be slow. This leads to an overall much higher segregation rate than the recovery rate, as indicated in Figure 5. Once segregated regions will thus remain segregated long after the charge-carriers have recombined. This proposition is corroborated by the experimentally observed acceleration of segregation for Br-rich samples (x = 0.9), by a factor of 6 compared to x = 0.2. The stronger local charge-carrier concentration in Br-rich samples leads to higher local vacancy and interstitial densities. From the phase distribution of the final segregated state, we estimate as a lower limit a 10-fold higher local charge carrier concentration in Br-rich samples compared to I-rich samples. However, we anticipate, that this difference is even larger in the initial stage of phase-segregation, when the I-rich domains are below the detection limit of XRD. Higher spatial inhomogeneity in the vacancy density in Br-rich samples can thus well rationalize the faster segregation in the Br-rich samples.

How this mechanism leads to gradual phase-segregation needs to be investigated further. We here like to formulate some hypothesis, which we hope can guide the development of more advanced models and experiments to be verified or disproven. In an initial step, charge carriers accumulate on statistically-appearing low-bandgap domains. In these 'hot-spots' the local charge-carrier density becomes very high and the phase-segregated state becomes energetically favored, in line with the *charge carrier localization models*. Almost pure I-rich domains in close proximity to Br-enriched domains are formed, corresponding to Stage 1 in **Figure 5**. Interdiffusion between I-rich and Br-rich domains leads to local re-equilibration and concentration gradients, especially when domains are de-excited and the driving force for phase-segregation disappears. Due to the relatively larger energetic depth of iodide-rich domains in very high bromide-content surrounding, charge carriers funneling into newly appearing low-bandgap domains caused by random fluctuations - 'hot spots' – is preferred and a new area being segregated. Continuation of this may lead to a spot-by-spot segregation. Further kinetic modelling should therefore include the interplay between inhomogeneous and dynamically changing segregation rates as well as continuous slow diffusion driven remixing and continuous appearance of new I-rich domains due to random fluctuations.

Our results highlight the importance of photo-induced dynamically changing material properties in metal halide perovskites. To assume constant defect concentrations during illumination falls short of reality in most cases. Dynamic defect concentrations have been reported plentiful and are present in all metal halide perovskites independent of the halide mixture.[54–59] The resultant dynamic ion mobility is underlying photo- and charge-carrier induced effects such as hysteresis, photo-enhancement and photo-bleaching, as well as the remarkable recovery and self-healing capabilities of metal halide perovskites. Mixed halide perovskites present a special model system in which dynamic changes in ion mobility determine the phase segregation-process and therefore become uniquely apparent and examinable.

## 4. CONCLUSION & OUTLOOK

While a multitude of diverse approaches have been developed to rationalize the photo-induced phase-segregation there is a discontinuity between thermodynamic models rationalizing the initial stage of phase-segregation and models phenomenologically describing the latter stage of phase-segregation and its kinetics. To reconcile both regimes, we propose that a dynamic model is needed, considering the accumulation of light-induced material perturbations during continuous irradiation.

Qualitative comparison of our experimental results with current thermodynamic models reveals that the phase-distribution established during phase-segregation as well as the stability of the samples is likely to be determined by the energy gain of charge-carriers localizing in randomly appearing I-rich domains and therefore initializing nanoscopic phase-segregation. Both, the charge-carrier localization model and the polaron model rely on the energy of the average charge-carrier distribution and static material properties. However, a quantitative comparison of the final equilibrium state reached during illumination, yields that the energy of the average charge-carrier density is not sufficient to rationalize the severity of the segregation. Our experimental data gives quantitative insight into the full range of phases forming during segregation, their kinetics as well as the simultaneous evolution of the emissive phase for a complete set of sample compositions. We are confident, that this unique set of structural, kinetic and optical data delivers a basis for the development of further, more sophisticated models, considering the dynamic nature of metal halide perovskites.

As the exact mechanism behind the suggested gradual segregation will have to be explored. Based on the experimental findings described herein, we would here like to propose several hypotheses, to guide further experimental and theoretical works:

- The localization of charge-carriers on low-energy sites leads to an inhomogeneous charge-carrier density and may lead to the preferential generation of iodide vacancies and interstitials
- Due to differences in local ionic defect concentrations, there are differences in the segregation and remixing rates in charge carrier enriched (excited) and charge carrier deprived (dark) domains
- Macroscopic phase-segregation occurs *spot-by-spot* through accumulative changes in the material
- In addition to drift, diffusion-driven ionic transport due to gradients in the iodide vacancy and interstitial concentrations need to be considered

## 5. EXPERIMENTAL DESIGN AND DATA ANALYSIS

*EXPERIMENTAL DESIGN*

To follow the emergence of iodide-rich domains and their population with charge-carriers, as well as changes in the microscopic sample composition during illumination, we measured X-ray diffraction (XRD) and photoluminescence (PL) signals simultaneously at the same area of interest. For this we designed a custom-built experimental chamber, shown in **Figure 1**a. For sample illumination and PL excitation we used a blue LED centered at 453 nm and calibrated to 0.1 W cm$^{-2}$ power density at the sample surface (1 Sun). The experimental setup was designed to maintain a controlled atmosphere and constant temperature during the experiment. In the experiments reported here, the sample was kept in a nitrogen atmosphere at 20°C, since the photo-segregation process has been shown to be atmosphere[32] and temperature dependent[18,60] (see SI Note 13) For the in-situ XRD measurements, the high-flux GaKα radiation of a liquid metal-jet X-ray source in combination with an area detector was used.[61,62] Due to the splitting of the GaKα line into GaKα$_1$ and GaKα$_2$, the measured XRD patterns of the homogeneous samples show a double peak for each reflection. (see SI Note 1, Figure S2).

*SAMPLES*

We here studied methylammonium lead bromide/iodide, MAPb(Br$_x$I$_{1-x}$)$_3$ samples. The average composition, $\bar{x}$, of the samples was defined as $\bar{x}$=[Br]/([I]+[Br]). Samples with the average sample compositions of $\bar{x}$ = 0, 0.1, 0.2, 0.4, 0.6, 0.8, 0.9 and 1 were studied. As the sample's phase-distribution and hence ionic homogeneity varies during the photo-segregation process, it is important to distinguish the local halide composition of a sub-phase, $x$, within the phase-distribution, $\Phi(x)$.

of the sample from the average halide composition of the sample, $\bar{x}$, assumed to be equivalent to the halide ratio in the precursor solutions used for depositing the samples. Samples were prepared according to procedures described previously and details can be found in the SI Note 1.[44,63]

*ANALYSIS PROCEDURES*

Exemplary PL and XRD measurement data acquired during illumination are shown in Figure 1c and 1b for the sample with $\bar{x}$ = 0.6. It shows the shift of the PL peak energy from 1.9 eV to 1.7 eV upon illumination as previously reported (Figure 1b).[1,2,15,27,64] Simultaneously recorded changes in the 200 XRD reflex of the cubic $Pm\bar{3}m$[65] phase can be seen: the diffraction peak, initially centered at 25.5°, broadens remarkably upon illumination of the sample. We here focus on the analysis of the 200 peak, but the behavior is also observed for other reflexes in the detection range as discussed in the SI (see SI Note 1, Figure S3 for all full patterns from 10˚ to 32˚). It must be noted, that both the XRD pattern and the PL spectra reach an equilibrium, where further changes are negligible. We consider this state to be the equilibrium phase-distribution under illumination. In the case of $\bar{x}$ = 0.6, this equilibrium is reached after 20 min, as indicated by a dashed line in Figure 1b, c.

During illumination of our samples, we checked that we predominantly observe the light-induced phase-segregation and not photo-degradation effects or the formation of other secondary phases by evaluating the center of mass (COM) of the XRD diffraction patterns, defined as:

$$COM = \int I_{XRD}(2\theta)\, 2\theta\, d\theta / \int I_{XRD}(2\theta) d\theta \qquad \text{(Equation 5)}$$

where $I_{XRD}(2\theta)$ is the XRD intensity at $2\theta$. Neither the *COM* nor the integrated area under a specific XRD-peak, $\int I_{XRD}(2\theta)$, changed during the duration of the illumination, as discussed in SI Note 2 and 3. This confirms that no non-perovskite secondary phases are formed, no material is lost (e.g. by evaporation or other degradation processes), and neither are amorphous phases formed at the time-scale of the experiment.

We developed an analysis procedure to translate both the PL peak position as well as the XRD diffraction pattern to the composition or compositional distribution, $\Phi(x)$, within the MAPb($Br_xI_{1-x}$)$_3$ samples. This enables us to directly compare, and contrast compositional information gathered from the PL and XRD signatures from the samples under investigation. To translate the experimentally detected PL peak energy to a specific composition, $x_{PL}$, a calibration curve relating the PL peak position to the halide composition was used, as discussed in SI (Figure S8).[1,20,42,66] The phase-distribution, $\Phi(x)$, is derived from the XRD patterns, $I_{XRD}(2\theta)$. For this, the continuous phase-distribution $\Phi(x)$ is approximated by a discrete function $\varphi_m$ with 21 different compositions $x_m$ indexed by *m = 1, 2, ..., 21*. A corresponding XRD pattern can then be calculated for the distribution $\varphi_m$ through a linear combination of reference diffraction patterns $I_{XRD_m}(2\theta)$ expected for samples with different compositions We constructed reference patterns, $I_{XRD_m}(2\theta)$, with $x_m$ ranging from 0 to 1 in steps of 0.05 assuming that the XRD peak shapes are equal to the instrumental response function (IRF) at the diffraction angles expected for the specific composition $2\theta(x_m)$. These reference patterns were used to calculate the linear combination coefficients $\varphi_m$:

$$I_{XRD}(2\theta) = \sum_{m=1}^{21} \varphi_m \, I_{XRD_m}(2\theta) \qquad \text{(Equation 6)}$$

where the coefficient $\varphi_m$ is the volume fraction of the sample in the phase with composition $x_m$. This procedure is assuming that the width of the diffraction peaks in the dark and under illumination is dominated by compositional inhomogeneity and neither strain nor size. We justify this assumption in SI Note 3 and 6, where we show from both Williamson-Hall analysis and the analysis of the peak shape that peak-broadening is not consistent with changes in the domain size nor strain in the samples under investigation.

To quantify and compare the extent of phase-segregation between samples, we introduce the *degree of segregation, $D_{seg}(\bar{x})$*. The degree of segregation measures how much the compositional distribution is different from that expected for an ideal homogeneous sample of composition $\bar{x}$ normalized to the state where the sample is segregated entirely into pure iodide and pure bromide domains (as illustrated in Figure S13). Thus, it is defined as the ratio between the variance of the distribution $Var(\Phi(x,\bar{x}))$ and the maximum possible variance, $Var(\Phi_{max,seg}(x,\bar{x}))$ for a given composition $\bar{x}$:

$$D_{seg}(\bar{x}) = \frac{Var(\Phi(x,\bar{x}))}{Var(\Phi_{\max seg}(x,\bar{x}))} \approx \frac{\sum_{n=1}^{21} \varphi(x_m)[x_m - \bar{x}]^2}{(1-\bar{x})\bar{x}^2 + \bar{x}(1-\bar{x})^2} = \frac{\sum_{n=1}^{21} \varphi(x_m)[x_m - \bar{x}]^2}{2\bar{x}(1-\bar{x})} \qquad \text{(Equation 7)}$$

The maximum variance for a composition $\bar{x}$, , $Var(\Phi_{max\,seg}(x,\bar{x}))$ is reached when the sample is completely segregated into the pure halide compounds MAPbI$_3$ and MAPbBr$_3$: $\Phi_{max\,seg}(0) = 1 - \bar{x}$, $\Phi_{max\,seg}(0 < x < 1) = 0$ and $\Phi_{max\,seg}(1) = \bar{x}$. The maximum relative amount of MAPbI$_3$ and MAPbBr$_3$ are defined by the average sample composition $\bar{x}$. As an example for the sample with x = 0.6, $Var(\Phi_{max\,seg}(x, 0.6))$ would be reached when the sample is fully segregated into MAPbI$_3$ and MAPbBr$_3$ with a 40% and 60% volume fraction, respectively. The degree of segregation, $D_{seg}(\bar{x})$, is thus a quantitative metric expressing the phase-distribution function $\Phi(x)$ as a single number. We found $D_{seg}(\bar{x})$ a useful metric to compare samples of different compositions and their temporal evolution of the phase-segregation, $D_{seg}(\bar{x},t)$ quantitatively.

**Supporting Information**

Supporting Information is available from the Wiley Online Library or from the author.

**Acknowledgements**

We thank Masaru Kuno for valuable discussion on the charge carrier localization model. We thank Hampus Näsström for discussion and a valuable reference.
E. U. and K. S. acknowledge financial support from the Swedish Research Council (grants no. 2015-00163 and 2018-05014) and Marie Sklodowska Curie Actions, Cofund, Project INCA 600398 and Nano Lund. E. L. U., A. M. and C. R. also like to acknowledge financial support from the German Federal Ministry of Education and Research (BMBF–NanoMatFutur Project HyPerFORME: 03XP0091). C. R. acknowledges financial support from the HI-SCORE research school of the Helmholtz Association. I.S. acknowledges support from Swedish Research Council grant 2020-03530. E.U. and J.J. acknowledge funding from the Swedish Foundation for Strategic Research (SSF, Project ITM17-0276).

Received: ((will be filled in by the editorial staff))
Revised: ((will be filled in by the editorial staff))
Published online: ((will be filled in by the editorial staff))

**Table of content entry:**

Photo- and charge-carrier induced ion migration is a major challenge when utilizing metal halide perovskite semiconductors. It is uniquely apparent during the light induced phase-segregation of mixed iodide/bromide perovskites. We here investigate its mechanism and cause by simultaneous in-situ X-ray diffraction and photoluminescence spectroscopy during illumination for the full compositional range of $MAPb(Br_xI_{1-x})_3$ with $x = 0…1$.

Klara Suchan, Justus Just, Pascal Becker, Carolin Rehermann, Aboma Merdasa, Roland Mainz, Ivan G. Scheblykin* and Eva Unger*

**Multi-Stage Phase-Segregation of Mixed Halide Perovskites under Illumination: A Quantitative Comparison of Experimental Observations and Thermodynamic Models.**

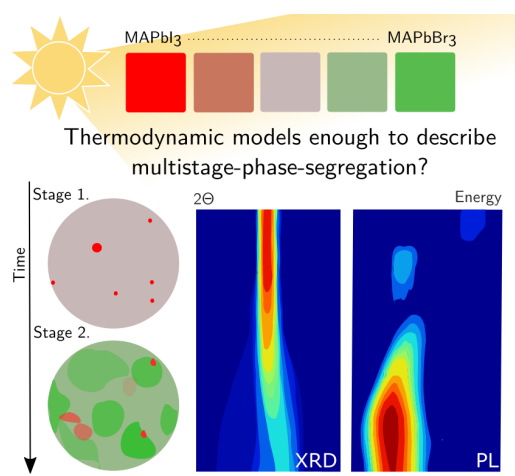

**Supporting Information (SI) for:**

**Multi-Stage Phase-Segregation of Mixed Halide Perovskites under Illumination: A Quantitative Comparison of Experimental Observations and Thermodynamic Models.**

*Klara Suchan[1], Justus Just[2], Pascal Becker[3,4], Carolin Rehermann[5], Aboma Merdasa[5], Roland Mainz[4], Ivan G. Scheblykin\*[1] and Eva L. Unger\*[1,5]*

[1]Division of Chemical Physics and NanoLund, Lund University, Box 124, 22100 Lund, Sweden, [2]MAX IV Laboratory, Lund University, PO Box 118, SE-22100 Lund, Sweden, [3]Helmholtz-Zentrum Berlin für Materialien und Energie GmbH, Structure and Dynamics of Energy Materials, Hahn-Meitner-Platz 1, 14109 Berlin, Germany, [4]Helmholtz-Zentrum Berlin für Materialien und Energie GmbH, Microstructure and Residual Stress Analysis, Albert-Einstein-Straße 15, D-12489 Berlin, Germany, [5]Helmholtz-Zentrum Berlin für Materialien und Energie GmbH, Young Investigator Group Hybrid Materials Formation and Scaling, HySPRINT Innovation Lab, Kekuléstraße 5, 12489 Berlin, Germany

*SI-Note 1:* Experimental methods

**Sample Fabrication**

Microscope glass substrates were cleaned in an ultrasonic bath in Mucasol, Aceton and Isopropanol (10 min in each solvent). Subsequently they were dried and cleaned for 15 min in O3-Plasma. Lead(II) iodide and Lead(II) bromide were purchased from TCI (Tokyo Chemical Industry UK Ltd.), MAI and MABr from Greatcell solar and anhydrous N,N-dimethylformamide (DMF) from Sigma Aldrich. All chemicals were used as received. Two solutions were prepared. The iodide-rich perovskite precursor solution was prepared by dissolving 159 mg of MAI and 461 mg of $PbI_2$ into 1 ml anhydrous N,N-dimethylformamide (DMF) . Analogously the bromide-rich perovskite precursor solution was prepared by dissolving 112 mg of MABr and 367 mg of $PbBr_2$ in 1 ml DMF to make 1 M solution with an equimolar ratio of precursor salts. To prepare the mixed halide sample series the two stock solutions were mixed in the appropriate ratios. 60μl of the mixed solution were then spin cast onto the substrates at 4000 *RPM* for 30*s*. After spin casting, the samples were annealed on a hotplate at 100 °C for 30 min. We studied $MAPb(Br_xI_{1-x})$ samples with compositions $\bar{x}$=0.1,0.2,0.4,0.6,0.8,0.9 and 1. To characterize a mixed halide perovskite, we introduce the term mean composition $\bar{x}$. Due to the phase segregation, the mean composition $\bar{x}$ may differ significantly from the nanoscale composition x.

**Measurements**

All measurements were performed in a constantly $N_2$ flushed enclosure (shown in Figure 1). The temperature was kept constant at $20 \pm 0.5°C$ using a Peltier element.

*Photoluminescence Spectroscopy* For in-situ photoluminescence measurements a custom setup was used in which the entire sample was excited with a blue LED (455 nm) at 100mW/cm² and the PL was detected with a fiber-coupled spectrometer as depicted in Figure 1. The such detected PL prior to and upon illumination for the sample $\bar{x}=0.6$ is shown in Figure S1.

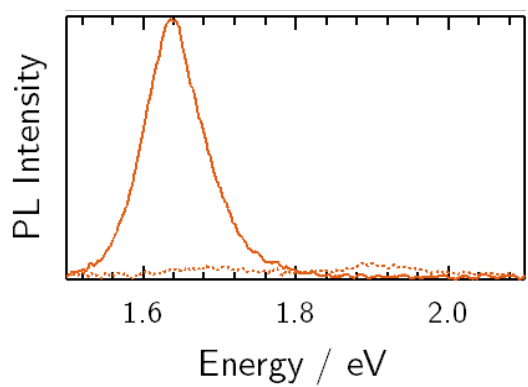

**Figure S1**: The Photoluminescence signal prior to (broken line) and upon illumination (solid line) of the sample $\bar{x}$=0.6. Excited with a blue LED at 455nm at 100mW/cm².

*X-ray Diffraction*

Simultaneous *in-situ* X-ray diffraction measurements were performed at the LIMAX laboratory[1] at BESSY II (Berlin), equipped with a liquid-metal-jet X-ray source, using GaK$_\alpha$-radiation. Please note that the GaK$_\alpha$-radiation consists of two lines, K$_{\alpha 1}$=9.25 keV and K$_{\alpha 2}$=9,22 keV, such that the instrumental response function consists of two closely situated lines. The instrumental response function can be assumed to be close to the pattern measured for a LaB$_6$ reference sample, shown in Figure S2b,c. The samples were measured with temporal resolution of 3 s. The size of the spot of the X-ray beam at the sample can be estimated to be approx.

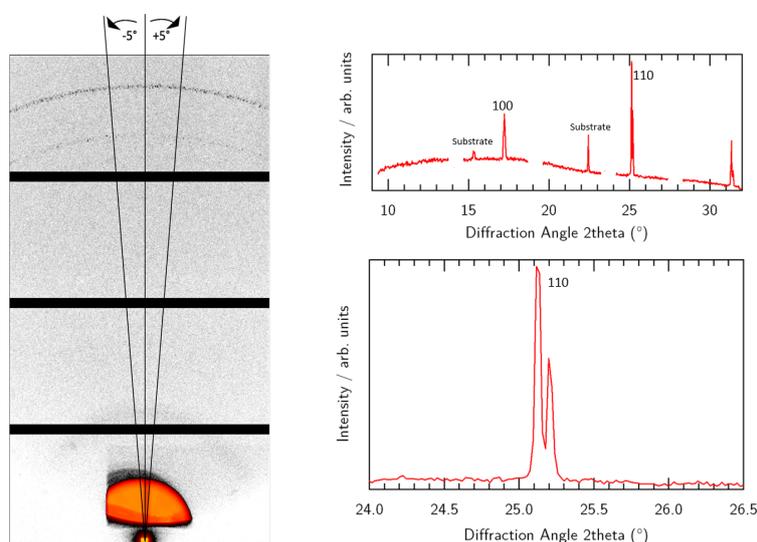

**Figure S2**: a) The raw x-ray diffraction pattern is shown here exemplarily as obtained from the 2D detector-array. It was subsequently integrated over an arc between +5° and -5° resulting in a pattern of intensity over diffraction vector length. Using a LaB$_6$ reference sample this was calibrated. b) The pattern for the LaB$_6$ reference from 9° to 32 °. c) is showing a close up to the 110 peak, revealing the double peak structure of the IRF.

5.5mm x 1mm. For the detection, an array of area detectors (PILATUS3 R 1M) was used. A range of 2θ from 10° to 32° was measured in Bragg-Brentano geometry with the focus on the 200 reflex (2θ ≈25°). Please not that there are gaps in between individual detectors (black horizontal bars in the raw x-ray pattern (Figure S2a)), even resulting in gaps in the pattern (Figure S2b). 1D XRD patterns were obtained by integration over an arc of the 2D diffraction patterns from 5 ° to -5 °azimuthal angle as can be seen in the Figure S2a. The full diffraction pattern of all compositions $\bar{x}$ can be seen in Figure S3.

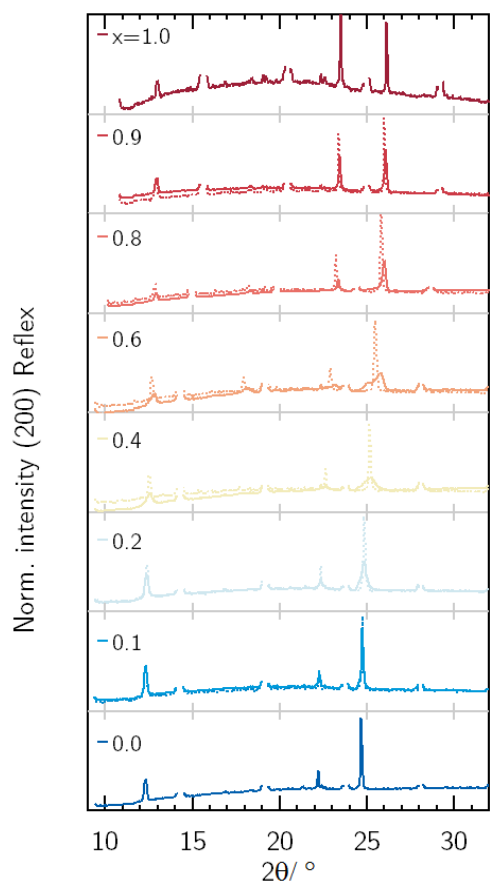

**Figure S3:** Full XRD Pattern from 2θ=10° to 32° for all compositions of MAPb(Br$_x$I$_{1-x}$)$_3$ with $\bar{x}$=0 to 1 before illumination (dashed lines) and after illumination (solid lines). The Gaps in the pattern are due to gaps in the detector array.

***SI Note 2:*** *Calculation of the arithmetic mean of all points on the x axis weighted by their individual intensity (center of mass)*

The center of mass of the XRD patterns can be used to investigate any change in net strain. Both compressive and tensile strain can occur in a material. Compressive strain leads to a shift of the XRD peak to higher angles, while tensile strain leads to a shift to lower angles. As compressive strain and tensile strain may compensate each other, the shift of the entire XRD peak corresponds to a change in the net strain in the sample. To calculate the peak position of the entire peak independent of peak shape, an arithmetic mean of all points on the horizontal axis weighted by their individual intensity can be calculated according to:

$$COM = \int XRD(2\theta)\, 2\theta\, d\theta / \int XRD(2\theta) d\theta$$

This will further be referred to as center of mass (COM) of the peak. The COM of the sample $\bar{x}$=0.8 before and after illumination can be seen in Figure S4. While in the unsegregated case the peak maximum agrees with the center of mass, in the photo segregated case, the peak position is shifted with respect to the COM. The Peak is thus asymmetric around the COM. Even if tensile and compressive strain compensate each other, such that the net strain is zero and the COM stays constant, the XRD peak may get affected by the strain. If some parts of the material are under compressive and others under tensile stress, part of the XRD peak is shifted to higher angles and another part of the peak to lower angles, leading to a broadening of the peak. Such a broadening due to strain is typically of Gaussian shape. We can thus conclude that the asymmetric peak shape with respect to COM as observed for the photo segregated sample with $\bar{x}$=0.8 cannot be explained by an increase of strain alone.

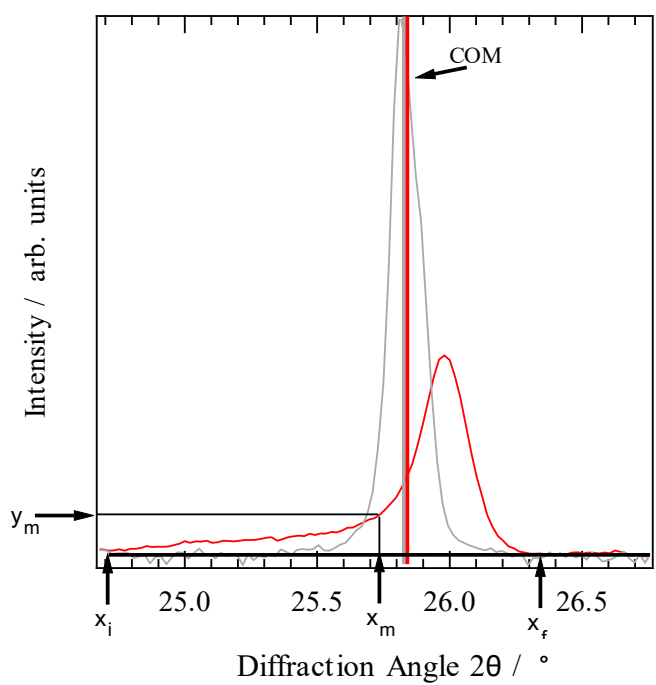

**Figure S4:** Diffraction pattern of the 200 reflex of sample $\bar{x} = 0.8$ before and after illumination(grey/red). The COM of each is represented by a solid vertical line in grey and red, respectively.

***SI Note 3*** *Width of the XRD Peaks*

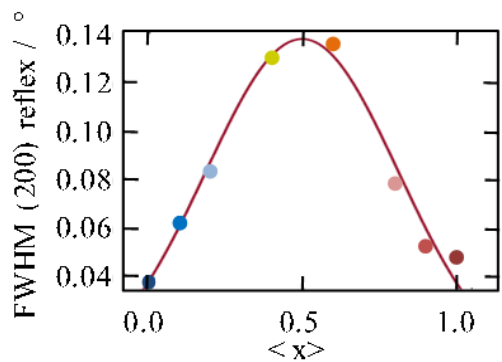

**Figure S5**: Comparison of the FWHM of the 200 reflex for different prior to illumination shown to follow a gaussian distribution as indicated by the fit (red line)

*Dark State*

Already before illumination the peak width is composition dependent following a gaussian distribution (Figure S5). The pure I or Br samples show quite narrow peaks while the width is increasing upon halogen substitution reaching the largest broadening at $\bar{x}= 0.5$. To assess the meaning of this composition dependent broadening, its origin needs to be evaluated.

A comparison of the Williamson-Hall analysis for $\bar{x}= 0$ and $\bar{x}= 0.6$ is used to evaluate the origin of this broadening. (Figure S6a) [2] The peak width B, times $\cos 2\theta$, for all visible peaks between $2\theta=10$ to $30$ is plotted over $\sin(2\theta)$. The pattern of an unstrained $LaB_6$ reference with average grain size of 10 µm is used for comparison. The peak broadening and thus the slope observed from the W.H. plot for the $LaB_6$ reference can be assumed to give an accurate measure of the instrumental response. The W.H. analysis can be used to detangle the broadening due to the crystallinity and the inhomogeneity. For the sample $\bar{x}=0$ only a very minimal difference to the $LaB_6$ reference can be seen. This proofs that the $\bar{x}=0$ is highly crystalline and nearly unstrained. The slight increase of the y-intercept from 0.16 to 0.17 for $\bar{x}=0.6$ compared to $\bar{x}=0$, indicates, that there is a slight decrease in the coherence volume, which is connected to the grain size and thus the crystallinity. However, if the entire peak broadening was caused by a reduction in the coherence volume, the curve for $\bar{x}=0.6$ would be expected to shift upwards but the slope would remain unchanged indicated by the dotted line. Instead, the W.H. analysis shows, that the slope for the sample $\bar{x}=0.6$ deviates significantly from that of the reference and $\bar{x}=0$. This indicates that the increased peak width originates primarily from a lattice distance inhomogeneity in the mixed crystal. This lattice distance inhomogeneity can either originate from an increase of strain or an increase of chemical inhomogeneity, such as an already existing phase segregation. Both are expected to increase for stoichiometries close to $\bar{x}=0.5$.

Previous reports show similar composition dependence of the peak width, a detailed comparison of the peak widths obtained here and literature values is included in SI Note 10.

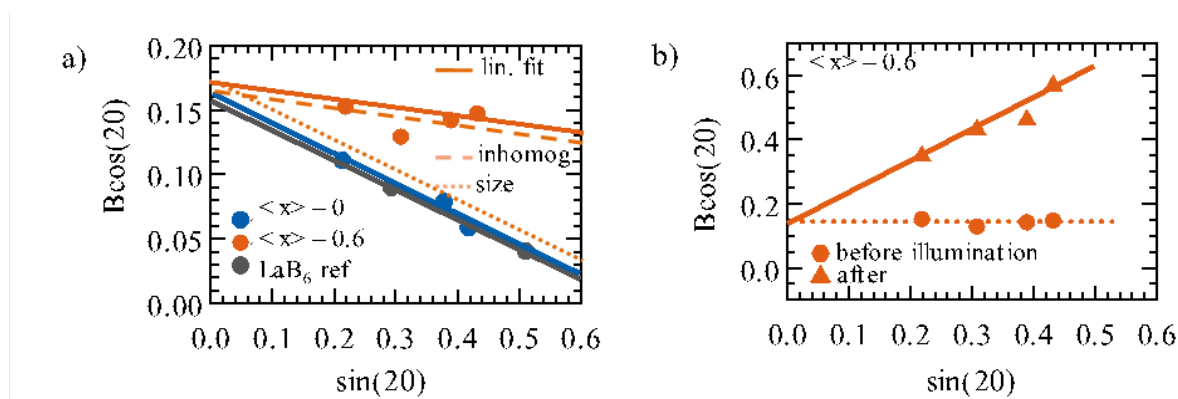

**Figure S6**: a) A Williamson-Hall analysis, where B is the peak broadening, is shown for samples with $\bar{x} = 0.6$ and $\bar{x} = 0$ before illumination and for a reference sample (LaB$_6$ powder). The dotted/dashed lines show the limiting cases for the peak broadening expected due to only a decreased size / only increased inhomogeneity, respectively. The main contribution to the increased peak broadening is due to an increase in the inhomogeneity of the lattice distances. d) Williamson-Hall analysis on the $\bar{x} = 0.6$ sample before and after illumination.

*Segregated equilibrium state*

Upon illumination, these mixed samples show an additional broadening, which has previously been assigned to an increase of the inhomogeneous micro strain.[2,3]. The Williamson-Hall analysis (Figure S6b) reveals that even under illumination, this broadening is caused by an increase in lattice distance inhomogeneity rather than a decrease of the domain size. This indicates that the light induced phase segregation does not result in a significant decrease of the crystallite size.

We would like to emphasize here, that any variation, δd, of the lattice plane distance, d, independent of its origin, will result in the same peak broadening, δθ, according to the Stokes Wilson Formula (Figure S7):

$$n\lambda = 2d\sin\theta \rightarrow n\lambda = 2(d - \delta d)\sin(\theta - \delta\theta) \rightarrow \delta\theta = -\frac{\delta d}{d}\tan\theta$$

From this equation, the peak broadening due to compositional inhomogeneity is virtually indistinguishable from the strain-related broadening.

Thus, a closer look at the peak shape can help to disentangle whether this inhomogeneity is strain related or due to compositional inhomogeneity:

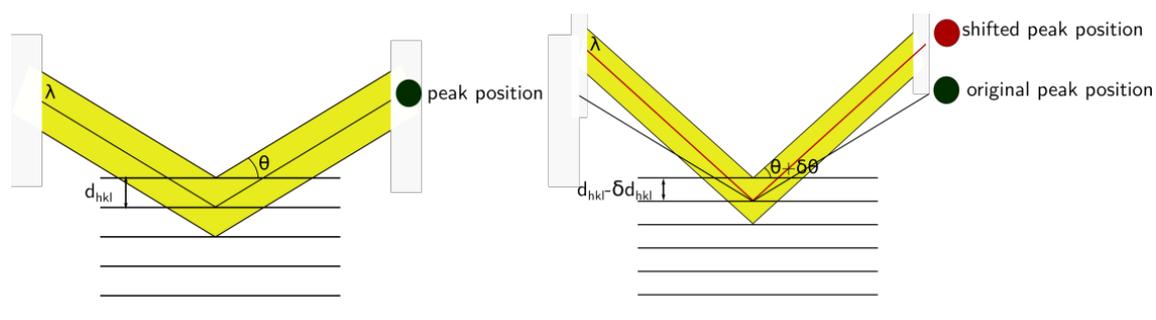

**Figure S7:** Sketch of the diffraction of an X-ray beam with wavelength λ on a crystal with lattice plane distance $d_{hkl}$ and a crystal with reduced lattice plane distance of $d_{hkl}-\delta_{dhkl}$, resulting in two different diffraction angles of θ and θ-δθ.

Micro strain commonly leads to a homogeneous (Gaussian) peak broadening. The strained peak should thus resemble a convolution of the original peak shape (Pseudo Voigt) and a Gaussian.[4] On the contrary, a broadening due to compositional inhomogeneity does not have this constraint. In the samples with Br-rich composition ($\bar{x} \geq 0.6$), illumination induces a peak shift accompanied by peak broadening and a long I-rich tail. The peak shift alone might look as if the sample is compressively strained. However, a more detailed look at all patterns shows, that even though the peak maximum is shifted significantly (e.g. in the composition $\bar{x}=0.8$ the peak shifts by 0.2 degrees), the center of mass (discussed in SI Note 2) of all patterns stays constant (Figure 2d). This is due to a broad distribution of minor scattering intensity at lower angles (24.5-25°). The constant centre of mass of the patterns shows that despite of the, for some compositions, large shift of the peak maximum, the net-strain in the sample stays constant during illumination. Any increase in compressive or tensile strain in the sample would thus have to be exactly compensated. This would lead to a symmetric Gaussian broadening around the position of the original peak (further discussed in SI Note 2).[4] The measured patterns, as shown in Figure 2a, however are asymmetric with respect to the original peak position. This allows for the assumption that during the phase segregation there is no relevant increase in strain.

***SI Note4:*** *Correlating Composition with PL and XRD Peak position*

In the absence of light, the bandgap, and with it the PL peak energy of MAPb($Br_xI_{1-x}$)$_3$ films, increases for higher Br content due to the smaller size of Br as compared to I as shown previously.[5,6] This increase is almost linear, being well described by modifying Vegard's law[7] by a small additional bowing factor, as shown in Figure S8b:

$$E_g(\bar{x}) = E_{g_{Br}} \cdot x + E_{g_I}(1-\bar{x}) - \bar{x} \cdot (1 - \bar{x}) \cdot b.$$

Assuming, that the PL stems from the band-to-band transition and as prepared samples are sufficiently homogeneous, this correlation between the nominal mean composition $\bar{x}$ and the PL peak energy can be used to estimate the composition of the local phase responsible for PL. We will here refer to such obtained compositions as $x_{PL}$. However, it has to be noted, that from PL measurements, we can obtain compositional information only about states/phases which contribute to the photon emission. Due to relaxation of charge carriers in an inhomogeneous material, these states/phases will contain the lowest part of the electron's density of states of the material. XRD is sensitive to the whole crystalline material in the sample and hence yields information about the overall distribution of compositions in the sample. In absence of light, an increase of the Br content results in an increase in the scattering angle of the mixed MAPb($Br_xI_{1-x}$)$_3$ films in XRD which can also be described by Vegard's law with an additional small bowing factor shown in Figure S8 a, in agreement with previous results[6,8]. Using the relation between scattering angle and Br-content ($\bar{x}$), the position of the XRD peak can be correlated to the materials composition.

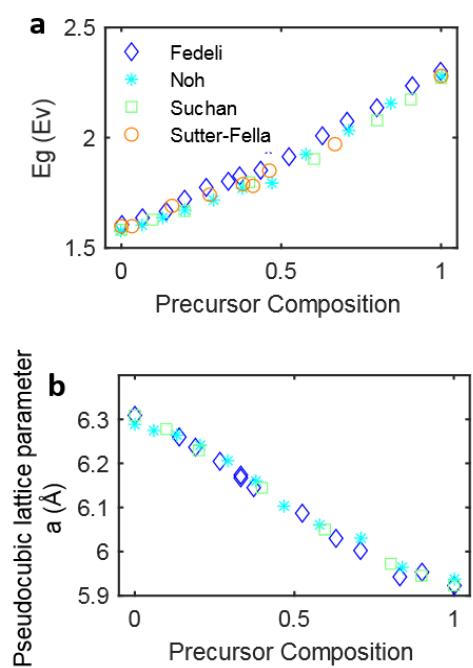

**Figure S8**: a) The photoluminescence peak position for all different compositions analyzed in this study is shown (green squares) and compared to the bandgap as reported previously b) Analogous, the pseudocubic lattice parameter is shown for all compositions (green squares) and is compared to literature values.

*SI Note 5: Differential X-ray diffraction pattern*

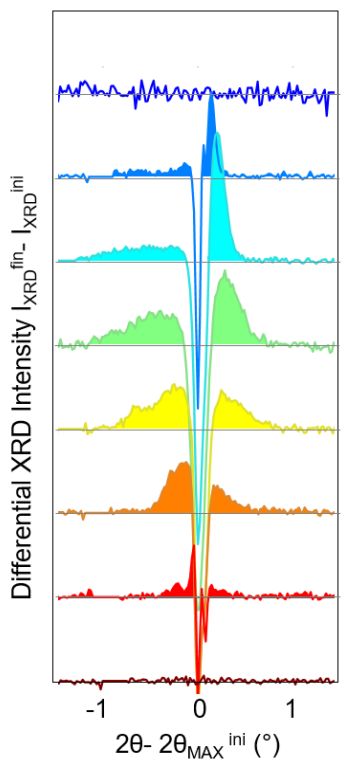

**Figure S9:** the differential XRD patterns show the change in the XRD intensity upon segregation. The long I-rich tail that establishes in Br-rich samples becomes clearly visible.

The changes in the XRD patterns can be uniquely illustrated by plotting the differential scattering signal, $I_{XRD_{fin}}(2\theta) - I_{XRD_{ini}}(2\theta)$, relative to the initial maximum peak position of the dark state, $2(\theta - \theta_{ini}^{max})$, shown in S9. This differential plot of the XRD patterns illustrates that the peak broadening upon illumination of Br-rich samples is asymmetrical relative to the peak position in the dark $\theta_{ini}^{max}$. For Br-rich compositions, the peak maximum in $I_{XRD_{fin}}(2\theta)$ is clearly shifted towards higher angles compared to $I_{XRD_{ini}}(2\theta)$ indicative that the majority of the sample becomes Br-enriched, as illustrated in section 2.1, Figure 2c. Meanwhile, the long tail towards low diffraction angles corresponding to a broad distribution of I-richer compositions is only visible for Br-rich compositions. I-rich samples do not exhibit a corresponding tail to higher angles indicating that phase-segregation due to ion migration in I-rich samples does not lead to the formation of a broad distribution of Br-enriched domains. By plotting the differential XRD patterns, it can be assured, that the tail is no measurement artefact or stemming from an incorrect background subtraction.

*SI Note 6:* *Fitting of the X-ray diffraction pattern:*

Before illumination, the 200 peak of the diffraction pattern for each composition can be fitted with a Pseudo Voigt (PSV) with a unique peak width, (Figure S8). The such obtained peak width is small for samples with compositions close to $\bar{x}=0$ and $\bar{x}=1$ and broadens for compositions close to $\bar{x}=0.5$ as (Figure S5). The phase segregated diffraction patterns are additionally broadened compared to the non-illuminated film, which can be seen in Figure 2a.

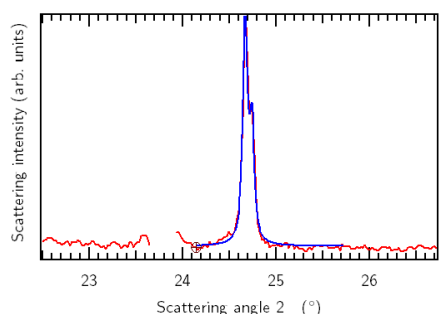

**Figure S10:** Fit of the X-ray diffraction pattern measured from the sample with composition $\bar{x}=0$ by a double Pseudo Voigt (PSV) function (describing the two diffraction peaks due to the two X-ray energies used $GaK\alpha_{1,2}$).

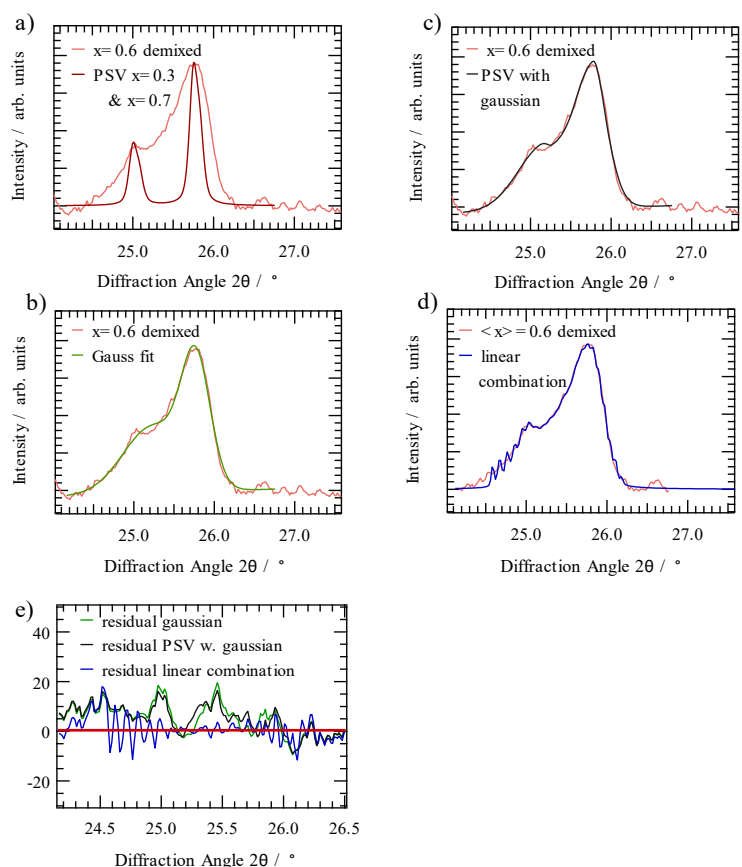

**Figure S11:** Fitting the X-ray diffraction pattern measured from the sample with composition $\bar{x}$=0.6. a) Fitting attempt with 2 Pseudo Voigt (PSV) peaks with the same peak width as found for the unperturbed samples. (Figure 2). The peak position resembles compositions of x=0.3 and 0.7. b) Fit with 2 PSV peaks with an additional gaussian broadening. c) Fit with 2 Gaussians with unrestricted peak width. d) Fit with a linear combination of 20 PSV peaks as described above. e) shows a comparison of the residuals for the fits shown in b,c and d.

In the case of $\bar{x}$=0.6, two peaks are visible in the diffraction pattern. In Figure S11a the 200 peak from the sample $\bar{x}$=0.6 is shown together with two PSV peaks of composition x=0.3 and x=0.7. The peak width of the PSV is kept the same as for the unilluminated samples of $\bar{x}$=0.3 and $\bar{x}$=0.7. This clearly illustrates that next to a peak split an additional broadening of each peak is occurring. If one assumes a split into two compositions and an additional peak broadening stemming from micro strain, the peak should resemble a convolution of the original peak shape (Pseudo Voigt) and a Gaussian. While this describes the high energy peak well, the fit of the low energy peak differs from the measured data (Figure S11 c). This results in a fitted average composition $\bar{x}$ as obtained from the fit of 0.7, which differs from the actual average composition of $\bar{x}$=0.6. Hence, assuming a split into two compositions with a strain related peak broadening does not accurately describe the pattern.

The pattern can neither be sufficiently described by a PSV function with an additional gaussian broadening, nor by Gaussian line shape (Figure S11 a,b,c)

Instead, all changes in the XRD patterns are consistent with phase segregation into a broad distribution of compositions x. If the broadening is induced by compositional inhomogeneity, the average composition $\bar{x}$ and thus the center of mass of the patterns must stay constant consistent with our observation. A variation of the local compositions within the film will lead to a multitude of narrow peaks at individual diffraction angles. The resultant diffraction pattern can hence be interpreted as a linear combination of XRD peaks from different (local) compositions. With this approach, both a peak shift and broadening as well as the peak

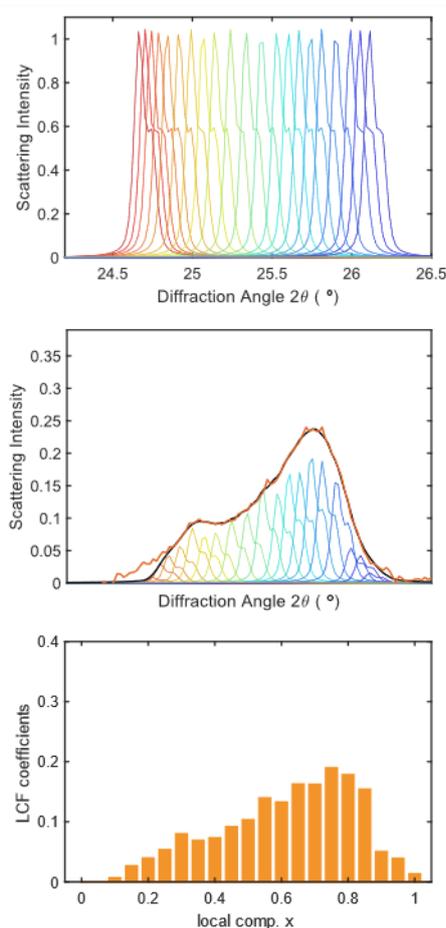

**Figure S12** a) The X-ray diffraction patterns of the 20 compositions used to fit the patterns of the phase segregated films with their linear combinations. b) The diffraction pattern of the phase segregated film of composition $\bar{x}$=0.6 (orange), the fit (black) and the contributing pattern (color according to Figure 2). c) The resulting linear combination coefficients are shown in a phase distribution histogram.

splitting can be described consistently. (Figure S12)

We chose to split the whole range of x from 0 to 1 into 21 compositions (Figure S12a) and to fit all XRD patterns with a linear combination of the 21 patterns corresponding to those compositions. This gives an exact solution by exploiting matrix operations:

c=AtA/AtB with c being the linear combination coefficients, A being a matrix of the 21 reference patterns and B being the experimental pattern to be fitted.

Figure S10b shows the resulting fit exemplary for $\bar{x}$ = 0.6. The coefficients of the linear combination of the 21 compositions provide a quantitative measure of the distribution of phases in the phase segregated sample (Figure S12c). The resulting estimation of the compositional distribution is shown for all samples investigated in Figure 3c.

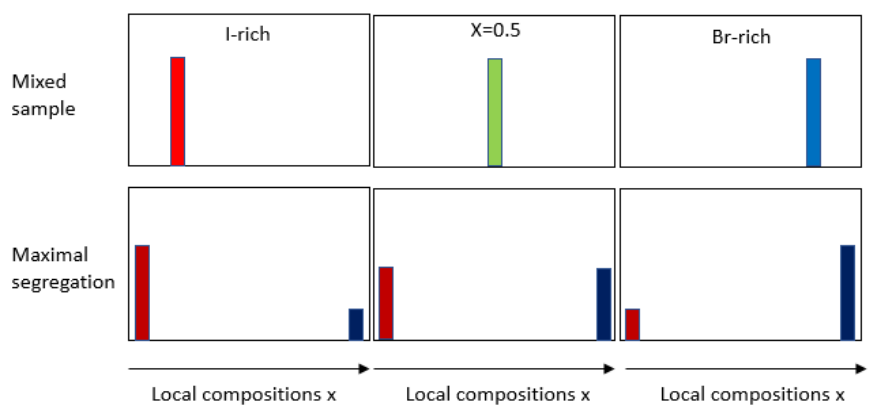

**Figure S13**: Scheme of the maximal segregation possible $D_{max}$ for I-rich and Br-rich samples and samples with compositions close to $\bar{x}$=0.5.

*SI Note 7*: *Degree of segregation for maximal segregation*
The degree of segregation gives an overall measure of how segregated a sample at any given time. The degree of segregation is normalized by the degree of segregation of a maximal segregated sample, which only consists of the pure phases, as discussed in section 5. In Figure S13 the maximal segregated case is shown schematically for an I-rich, a Br-rich and a $\bar{x}$=0.5 sample.

*SI Note 8: Charge carrier distribution*
While the conduction band is nearly isoenergetic, offsets up to 0.26 to 0.67 eV between the valence bands have been reported depending on the halide composition.[9–11] The formation of I-rich domains lowers the valence band energy locally, which generates a domain likely to trap holes.[7]

A rough estimation of the resulting charge carrier density in I-rich domains within the segregated film can be made by looking at the volume fraction of the emissive phase. The fraction of such low-bandgap sites decreases with increasing bromide content. For the $\bar{x} = 0.8$ sample, $x_{PL} \approx 0.1$. From the XRD pattern we derive that the volume fraction of that domain with x=0.1 is $\varphi_{x=0.1} = 1\%$.

The relative charge carrier density $n_x$, in these domains can be obtained by assuming, that the fraction of charge carriers that funnel there, $N_x$ is nearly 1. This is a justified assumption, as the high energy band in PL is quenched entirely during photo-segregation, signifying a depletion of charge carriers in the Br-rich domains. Hence the charge carrier density $n = N_{x=0.1} / \varphi_{x=0.1} \approx 1/0.01 = 100$.

Thus, the charge carrier density increases by a factor of 100 in I-rich domains in a $\bar{x} = 0.8$ sample. As the volume fraction of the emissive phase is decreasing with higher Br-content, the local charge carrier density is increasing with Br-content. This means, that the charge carrier density increases by a factor of 10 for $\bar{x} = 0.2$ but by a factor of 100 for $\bar{x} = 0.8$. For $\bar{x} = 0.8$ an illumination with 1 sun, thus results in a local charge carrier density within the I-rich phases in the segregated state comparable to an excitation with 100 sun in the mixed state.

*SI Note 9:* *Correlation of the evolution of PL-intensity with XRD kinetics*
*SI Note 10:* *Calculation of charge carrier density*

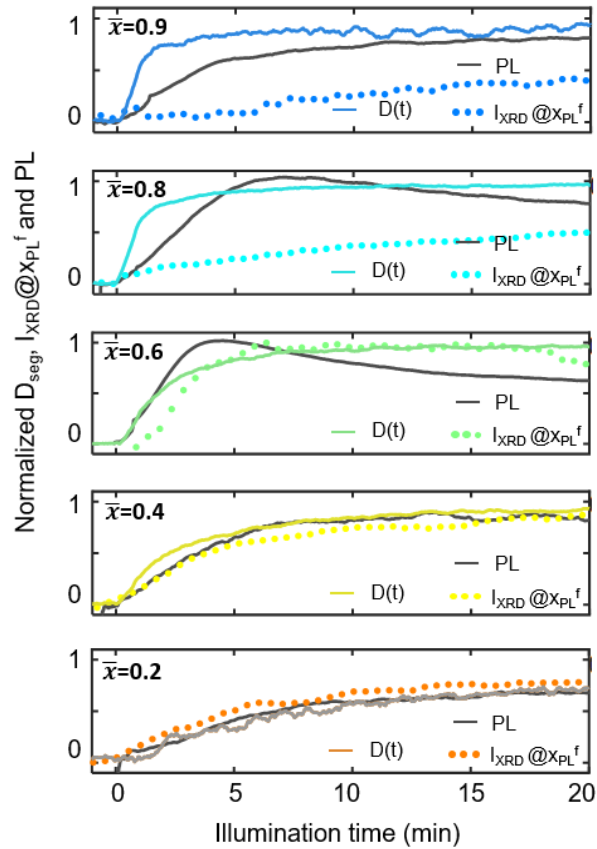

**Figure S14:** The segregation of the samples is shown to result in an increased PL intensity. This correlation is especially apparent for samples with $\bar{x} \leq 0.4$.

The charge carrier density N can be estimated from the excitation power. The excitation density was calibrated to be I=0.1W/cm2. With the photon energy of hv=4.3E-19 J (450 nm), the photon flux, $\Phi_{PH}$ can be calculated:

$$\Phi_{PH} = \frac{0.1 W/cm^2}{4.3\,E-19\,J} = 2.94\text{E}+17 \text{ photons} \cdot cm^{-2} \cdot s^{-1}$$

For the calculations, we consider a grain within the sample with an area of 300nm*300nm and a 200 nm thick film. A 200 nm think film can be assumed to be optically thick, such that it absorbs all photons at the wavelength of 450 nm.

With a charge carrier lifetime of $\tau = 100\ ns$, the number of generated carriers per cm² is:

$$n_{2D} = \tau * \Phi_{PH} = 2.94 \cdot 10^{10}\ carriers \cdot nm^{-2}$$

In an area of S=300 x 300 nm, this makes 25 photons, which are on average in the area.

Assuming the film thickness of 200 nm, the charge carrier density thus becomes

$n_{3D} = 1.5 \cdot 10^{-6} \; carriers \cdot nm^{-3}$

***SI Note 11:*** *Influence of charge carrier localisation and charge carrier lifetime on the energy gained by segregation*

The concentration of the 25 charge-carriers generated in a 300 nm x 300 nm x 200 nm domain into a small subdomain would lead to a local increased charge carrier density. As discussed in section 3.2, if those charge carriers would be concentrated on a 5 nm x 5 nm x 5 nm subdomain, this corresponds to a local enhanced charge carrier concentration by a factor of 25 000, their energy-density would be large enough to cause local phase segregation.

As the charge carrier density is proportional to the charge carrier lifetime, $n_{3D} \propto \tau$, an increased lifetime would enhance the charge carrier concentration within the entire sample. Thus, if $\tau$ was increased by a factor of 25 000, from 100 ns to 2.5 ms, phase segregation would not only be favored locally but in the entire sample. This can be seen in Figure S15. However, as 2.5 ms is an unrealistic long lifetime, we propose that charge carrier localization must play a significant role in phase segregation.

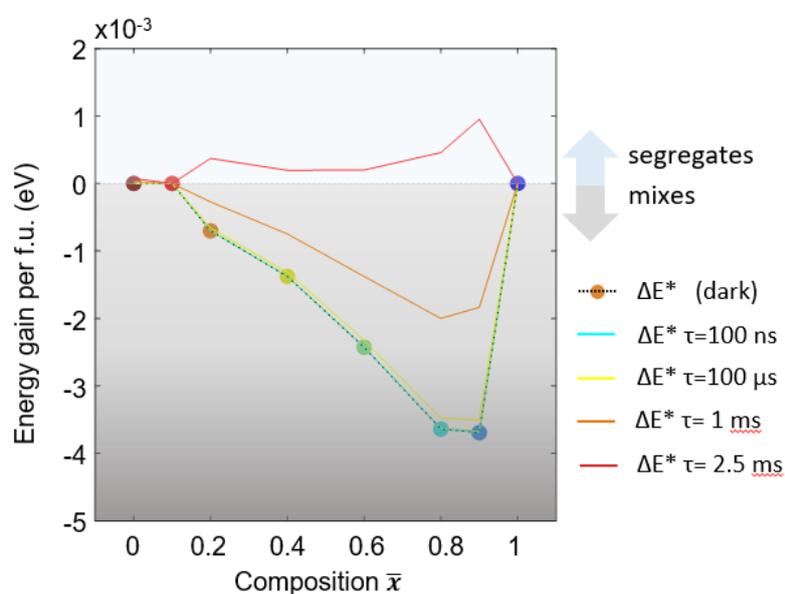

**Figure S15:** Energy gain by segregation considering different charge-carrier lifetimes.

***SI Note 12:*** *Sample Quality*

UV-Vis data of samples investigated in this study show the continuous change in bandgap upon varying the halide content of the sample (Figure S16). The flattening of the absorption at higher wavelength indicates the non-unity coverage of these samples. The absorptivity of samples at the excitation wavelength of the LED at 455 nm can be estimate from this data.

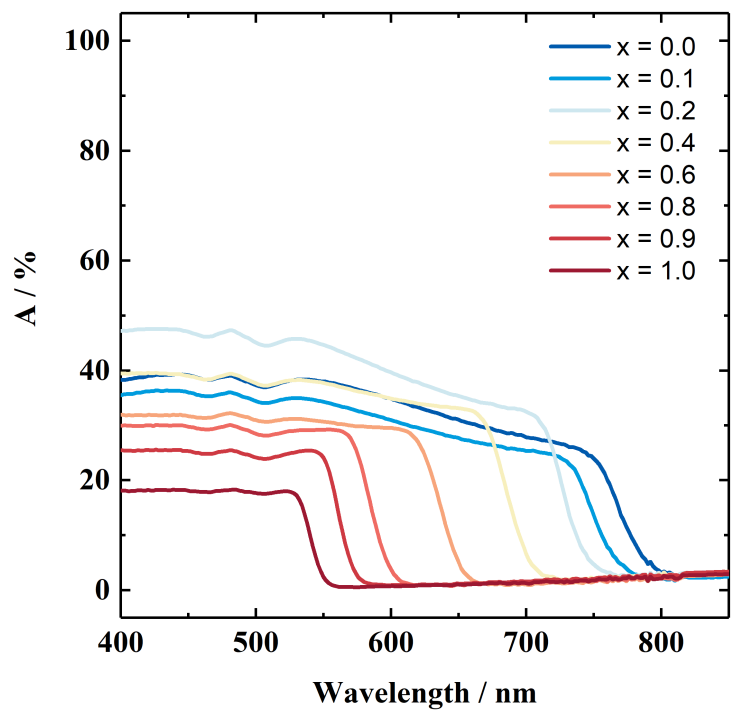

**Figure S16:** UV-vis measurements for all compositions of MAPb(Br$_x$I$_{1-x}$)$_3$ with $\bar{x}$=0 to 1.

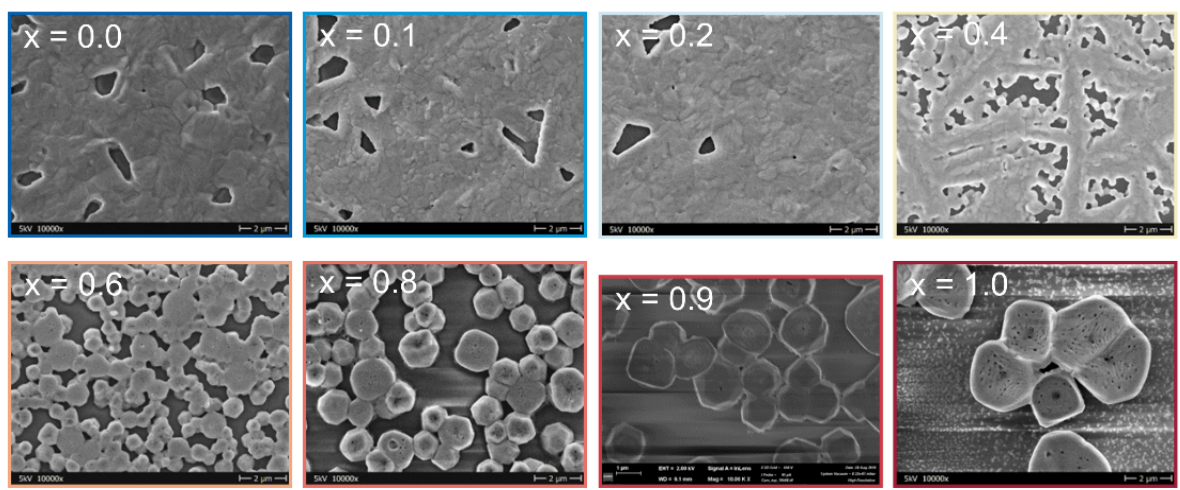

**Figure S17:** SEM top view images for all compositions of MAPb(Br$_x$I$_{1-x}$)$_3$ with $\bar{x}$=0 to 1 indicating different and non-unity coverage.

A comparison of the full width at half maximum (FWHM) of the 200 (cubic) or 220/004 (tetragonal) peak in XRD patterns of MAPb(Br$_x$I$_{1-x}$)$_3$ samples with varying halide content prior to illumination is shown in Figure S18. As reported previously, the samples analysed in this study exhibit a broader FWHM for intermediate halide compositions. This is usually

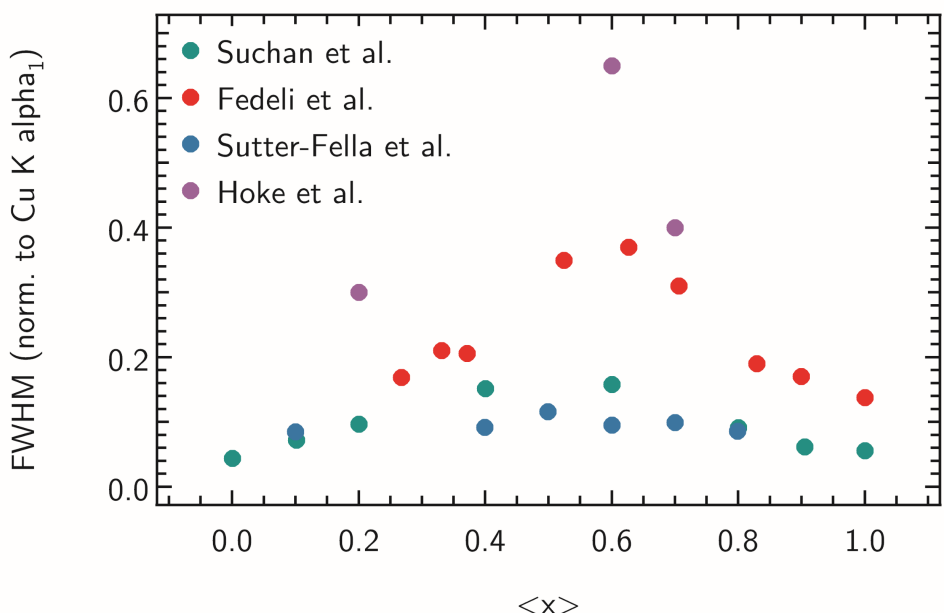

**Figure S18:** Comparison of the FWHM of the 200 (cubic) or 220/004 (tetragonal) peak in XRD prior to illumination as measured by us (green) and as reported in literature. All values are normalized to CuK$\alpha_1$ for better comparability

interpreted to be indicating some level of intrinsic ionic inhomogeneity in the samples as prepared but may also be due to increased strain within the sample due to the different sizes of iodide and bromide ions. The samples investigated here were pre-heated and cooled down, prior to the experiment to minimize any pre-existing phase segregation. They indeed show a relatively low FWHM for all compositions even for samples with equal amount of bromide and iodide, comparable to the results reported by Sutter-Fella et al. The herein observed samples can thus be assumed to be of comparably high crystallinity and low pre-existing halide segregation. Please note that the peak broadening in this study has a symmetric gaussian dependency on the composition. We thus conclude that if pre-existing compositional inhomogeneity remains even after heating the sample, this is unlikely to result in the observed asymmetry in phase segregation. In previous reports even an asymmetric increase of the peak-width, showing maximal FWHM broadening of XRD peaks at higher Br compositions has been observed (Fedeli et al.[12] and Hoke et al.[3]). This is likely due to more pronounced pre-existing phase-segregation[12,13] which we here avoided by heating samples to 90 °C and subsequently cooling it down to 20°C in dark conditions immediately before the measurement.

**SI Note 13:** *Reversibility, Reproducibility and Atmospheric conditions*

To ensure that no lasting structural changes were induced by exposure to light a sample with composition $\bar{x}$=0.6 was exposed to 1 sun multiple times as shown in Figure S19. XRD Pattern were recorded both, during the light induced phase segregation and during the subsequent remixing process in the dark. For the remixing process the sample was additionally heated to 90°C to ensure full mixing. Figure 19 shows, that the original peak height can be fully restored in the dark at elevated temperature. Thus, we assume that neither a loss of material occurred, nor a permanent change in the structure or crystal size.

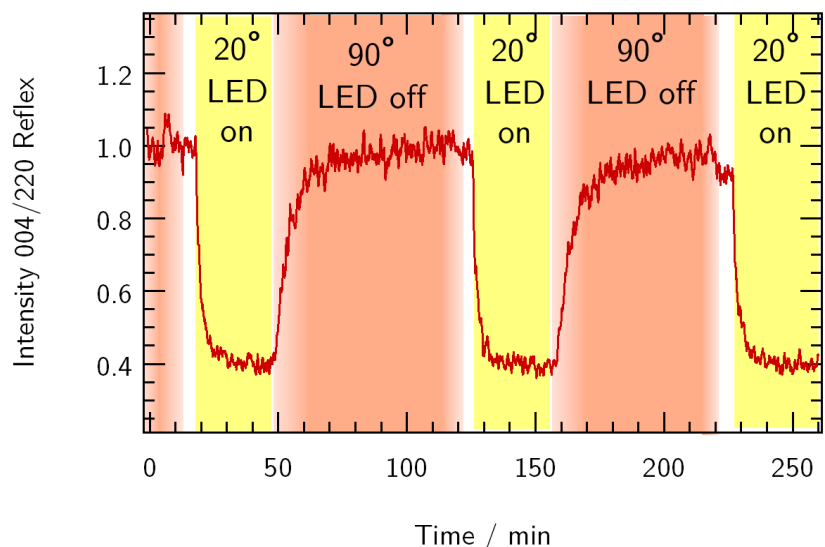

**Figure S19:** Intensity of the 004/220 reflex in XRD during multiple cycles of phase segregation and recovery at elevated temperature.

In order to investigate the degree of reproducibility of the compositional changes during phase segregation, the process was repeated several times for samples with $\bar{x}$=0.6. This composition is one which is very prone to segregate, resulting in a broad compositional distribution and distinct peak splitting into two dominant phases and thus was deemed the most challenging composition regarding reproducibility.

In Figure S20 it can be seen that prior to illumination the different samples of composition $\bar{x}$=0.6 show similar but slightly varying peak position and FWHM. This may be due to a dependence of the crystallinity to the precise processing conditions.

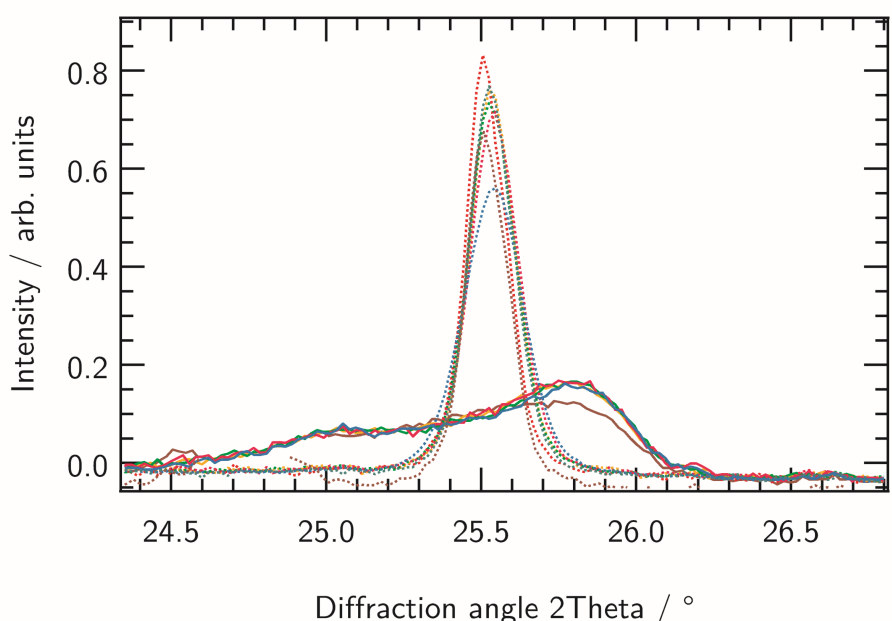

**Figure S20:** Comparison of the pattern of the 200 peak in XRD prior to (broken lines) and upon illumination (solid lines) for different samples of the composition $\bar{x}=0.6$

However, after phase segregation, the pattern for all samples are identical, indicating that the observed structural changes are rather fundamental and not heavily dependent on the precise original crystallinity or homogeneity and with this the processing conditions.

Furthermore, the influence of the atmospheric conditions on the phase segregation was analyzed. For this, in-situ PL spectra were recorded for different samples in N2, in air and covered in PMMA. The normalized colourmaps are shown in Figure S 21. It has to be noted, that the PL signal has been shown to be very sensitive to the atmospheric conditions, even for pure I perovskites. It can therefore be assumed, that the structural properties of will be less sensitive to the atmospheric conditions.

All the sample show qualitative similar behavior including a first peak around 650 nm which decreases in intensity within the first seconds. This is followed by a short lived intermediate peak above 700 nm, as reported previously[14]. Only after this the final PL peak is growing in red shifting from 710 to 750 nm. Slight differences in the PL intensity can be observed. While in N2 and in air, the initial and intermediate peak are very weak in intensity, the sample covered with PMMA shows a higher PL intensity already in the beginning, which could hint to passivation effects. The growth kinetic of the final PL peak is very similar for both N2 and the PMMA covered and a faster only for the unencapsulated sample in air. This could be due to an influence of oxygen on the PL yield.

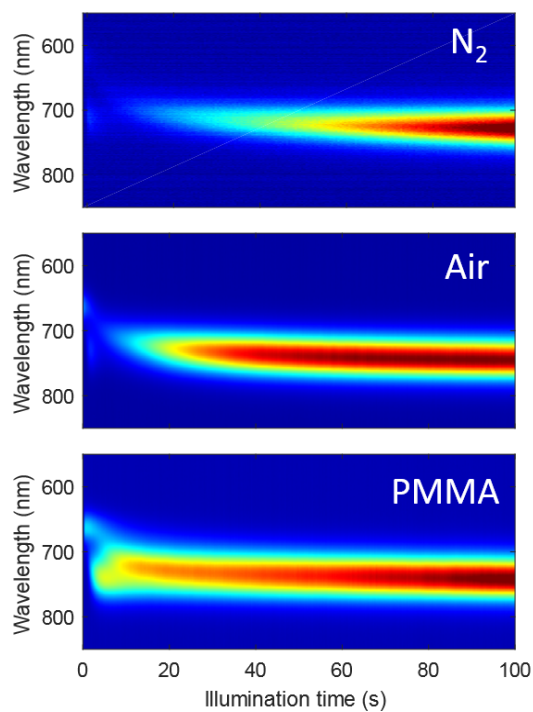

**Figure S21:** Colourmaps of the PL signal recorded during the light induced phase segregation of a samples with composition $\bar{x}=0.5$ in N2, air unencapsulated and covered with PMMA.

However, we conclude that the PL signal during phase segregation is rather weakly influenced by the atmospheric conditions.

(1) Wansleben, M.; Zech, C.; Streeck, C.; Weser, J.; Genzel, C.; Beckhoff, B.; Mainz, R. Photon Flux Determination of a Liquid-Metal Jet X-Ray Source by Means of Photon Scattering. *J. Anal. At. Spectrom.* **2019**. https://doi.org/10.1039/C9JA00127A.

(2) Williamson, G. K.; Hall, W. H. X-Ray Line Broadening from Filed Aluminium and Wolfram. *Acta Metallurgica* **1953**, *1* (1), 22–31. https://doi.org/10.1016/0001-6160(53)90006-6.

(3) Hoke, E. T.; Slotcavage, D. J.; Dohner, E. R.; Bowring, A. R.; Karunadasa, H. I.; McGehee, M. D. Reversible Photo-Induced Trap Formation in Mixed-Halide Hybrid Perovskites for Photovoltaics. *Chem. Sci.* **2014**, *6* (1), 613–617. https://doi.org/10.1039/C4SC03141E.

(4) de Keijser, T. H.; Langford, J. I.; Mittemeijer, E. J.; Vogels, A. B. P. Use of the Voigt Function in a Single-Line Method for the Analysis of X-Ray Diffraction Line Broadening. *J Appl Cryst* **1982**, *15* (3), 308–314. https://doi.org/10.1107/S0021889882012035.

Halide Perovskite. *Journal of Luminescence* **2020**, *221*, 117073. https://doi.org/10.1016/j.jlumin.2020.117073.